\documentclass[preprint,12pt,authoryear]{elsarticle}

\usepackage{natbib}
\usepackage{hyperref}
\usepackage{subcaption}

\usepackage{amssymb}
\usepackage{amsmath}
\usepackage{amsthm}
\usepackage{mathtools}
\usepackage{multirow}
\usepackage{multicol}
\usepackage{booktabs}
\usepackage{bm}
\usepackage{graphicx}
\usepackage{mathrsfs}
\usepackage{tabularx}
\usepackage{array} 
\usepackage{booktabs}
\usepackage{caption}
\usepackage[table,xcdraw]{xcolor}
\usepackage[normalem]{ulem}
\useunder{\uline}{\ul}{}
\usepackage{setspace}
\usepackage{lineno}

\journal{Computers \& Chemical Engineering}

\begin{document}

\begin{frontmatter}



\title{Leveraging Graph Neural Networks and Multi-Agent Reinforcement Learning for Inventory Control in Supply Chains} 


\author{Niki Kotecha, Antonio del Rio Chanona} 

\affiliation{organization={Imperial College London},
            addressline={Department of Chemical Engineering}, 
            city={London},
            postcode={SW7 2AZ}}

\begin{abstract}

Inventory control in modern supply chains has attracted significant attention due to the increasing number of disruptive shocks and the challenges posed by complex dynamics, uncertainties, and limited collaboration. Traditional methods, which often rely on static parameters, struggle to adapt to changing environments. This paper proposes a Multi-Agent Reinforcement Learning (MARL) framework with Graph Neural Networks (GNNs) for state representation to address these limitations.

Our approach redefines the action space by parameterizing heuristic inventory control policies, into an adaptive, continuous form where parameters dynamically adjust based on system conditions and avoid combinatorial explosion typical of discrete actions. By leveraging the inherent graph structure of supply chains, our framework enables agents to learn the system's topology, and we employ a centralized learning, decentralized execution scheme that allows agents to learn collaboratively while overcoming information-sharing constraints. Additionally, we incorporate global mean pooling and regularization techniques to enhance performance.

We test the capabilities of our proposed approach on four different supply chain configurations and conduct a sensitivity analysis. This work paves the way for utilizing MARL-GNN frameworks to improve inventory management in complex, decentralized supply chain environments.

\end{abstract}


\end{frontmatter}


\section{Introduction}
Modern supply chains are complex and operate under uncertain environments. These uncertainties can lead to disruptions and sub-optimal performance, often due to operational failures or poor coordination between different parts of the supply chain. The inventory control problem, a sequential decision-making problem, is additionally challenged by stochastic and volatile factors such as lead times and seasonal demand patterns, often resulting in sub-optimal performance. The impact of disruptions such as the bullwhip (demand amplification) or ripple effect (disruption propagation) effect can be alleviated through collaborative efforts among different entities within a supply chain \citep{de2015mitigation}.

\subsection{Inventory Control}

The theory of inventory control can be traced back to the news-vendor problem \citep{edgeworth1888mathematical, clark1960optimal}, and the first widely used numerical solutions for inventory optimization seems to be the Economic Order Quantity (EOQ) model \citep{erlenkotter1990ford}. These works were fundamental to the widely known policies in inventory control theory: (R,S), (s,S), (R,s,S) and (R,Q) \citep{heuristic1, heuristic2}. Over the years, control theory has had an influence on production planning and inventory control \citep{heuristic3, heuristic4}. 

One approach to find optimal inventory control policies is using traditional heuristics. These are simple, exact methods which suit relatively simple problems.  Methods such as reorder point heuristics \citep{heuristic1, heuristic2} and the Economic Order Quantity model \citep{erlenkotter1990ford}, provide exact solutions for structured settings. However, they lack adaptability, as they depend on pre-defined parameters that may not be effective in dynamic environments with fluctuating demand or unforeseen circumstances.

The complexity of modern systems has driven the adoption of dynamic programming, which enables the formulations of optimal policies by considering future states. However, obtaining exact analytical solutions often proves infeasible due to computational demands, especially in large-scale or highly stochastic environments.\citep{bellman1952theory}. This limitation has paved the way for advanced numerical and optimization-based techniques beyond traditional closed-form solutions. For example, optimization techniques like  Linear Programming (LP) \citep{janssens2011linear} or Mixed-Integer Linear Programming (MILP) \citep{you2008mixed} are applied to inventory management problems. LP is well-suited for problems with linear relationships, while MILP offers more flexibility by incorporating integer variables to address specific complexities like minimum order quantities. However, these methods do not directly account for uncertainty, making them less effective in highly stochastic environments.

To improve adaptability in dynamic settings, feedback-based control strategies such as Model Predictive Control (MPC) have been employed. MPC uses a rolling-horizon optimization framework, incorporating real-time system updates to enhance decision-making. Unlike static optimization, MPC continuously updates decisions based on new information,  making it particularly useful for inventory systems with demand variability \citep{dong2012novel}. One of the key advantages of MPC is its capacity to optimize control actions by predicting future system behavior over a finite horizon. This predictive capability enables the formulation of control strategies that can respond to changes in system dynamics, thereby improving decision-making processes in inventory management \citep{liu2011inventory}. For instance, \citet{liu2011inventory} demonstrated the application of MPC in optimizing production and distribution systems, highlighting its effectiveness in handling dynamic inventory challenges. Furthermore, the integration of feedback mechanisms allows MPC to adjust its predictions based on actual system performance, thereby reducing the impact of demand variability \citep{conte2005inventory}. However, its computational complexity scales poorly and it does not explicitly account for uncertainty, limiting its feasibility in large, stochastic environments \citep{ghaemi2007model}.

In contrast, stochastic \citep{stochastic} and distributionally-robust optimization approaches \citep{distributionalr1, distributionalr2}, offer ways to handle uncertainty explicitly. However, they face challenges such as online tractability, limiting their practical applicability in real-world supply chain scenarios. If the optimal policy is analytically intractable, methods like stochastic optimization can be used to find an approximate optimal policy \citep{grossmann2016recent, you2011stochastic}. While stochastic optimization often finds good solutions, it does not guarantee optimality. In specific scenarios, exact numerical methods like dynamic programming might be applicable \citep{berovic2004application, perez2021algorithmic}. However, these methods are often limited by scale due to computational demands.  This leads to approximate numerical methods. These methods provide scalability but may be at a sub-optimal performance. Additionally, Approximate Dynamic Programming (ADP) is a powerful technique specifically designed for problems with both dynamics (e.g., decisions impact future states) and stochasticity (e.g., uncertain outcomes) \citep{katanyukul2011approximate}. ADP leverages approximations and sampling techniques to find good solutions for complex inventory management problems that might be intractable for traditional methods. Reinforcement Learning methods, closely related to dynamic programming, also belong to the approximate dynamic programming methods class.

\subsection{Reinforcement learning for Inventory Control}

Reinforcement learning (RL) has emerged as a promising alternative for addressing the challenges of stochastic sequential decision-making problems. Its ability to excel in complex, dynamic environments and handle uncertainty, places RL as a valuable tool to enhance decision making in supply chains. While closely linked to dynamic programming, RL offers a general solution to identify approximate optimal policies for stochastic processes by leveraging the Bellman optimality equation to iteratively update value functions and improve decision-making policies over time. RL also provides a cost-effective solution for decision-making by enabling offline training, which reduces the online computational overhead compared to optimization approaches that require continual updates in receding or shrinking horizon frameworks.

Single-Agent RL methodologies have been explored extensively in this context to solve a Markov Decision Process (MDP), the mathematical framework used to model decision making in an inventory management system. Unlike traditional heuristic approaches that rely on pre-defined rules, RL leverages interaction with the environment to learn optimal policies. Reinforcement learning, particularly deep reinforcement learning, leverages neural networks to approximate the value functions and policies, enabling it to handle high-dimensional state and action spaces. Methods like Q learning \citep{slr3, srl4, srl6, srl15} and Policy Gradient methods \citep{srl14, burtea2024constrained, rangel2023application, rangel2024recurrent, shin2019reinforcement, yoo2021reinforcement}have shown promise in developing adaptive and scalable inventory policies that can learn from interactions with the environment over time. However, RL often struggles with integer or mixed-integer decisions, which are common in inventory management problems (e.g., order quantities). As the scale of inventory systems increases, the complexity of the decision-making process grows, resulting in a significantly larger action space. This increased complexity can hinder the effectiveness of RL algorithms, as they require extensive training data and longer convergence times to identify optimal policies. 

To address these challenges, recent work has explored action parametrization methods that enable RL to manage complex or hierarchical action spaces by combining continuous and discrete decision variables \citep{fan2019hybrid, bester2019multi}.  In the context of multi-agent control and decision-making, \citet{fan2019hybrid} introduced a hybrid actor-critic reinforcement learning model that operates in a parameterized action space, allowing the agent to handle both discrete and continuous action variables simultaneously. Their approach was demonstrated in RL environments such as Catching Point, Moving, Chase and Attack, and Half Field Football. \citet{khamassi2016active} and \citet{zhang2024model} introduced Parameterized Action Space Markov Decision Processes (PAMDPs) in the context of robotics and other RL environments, which demonstrated how continuous action spaces can be effectively utilized in complex manipulation tasks. This is relevant in inventory management as decisions often involve selecting order quantities which can be treated as discrete actions with continuous parameters. However, current studies in inventory control often represent the act of order replenishment as a discrete integer decision \citet{slr3, srl13}, rather than leveraging on pre-defined heuristic parameters like reorder point (s) or order-up-to level (S).. Notably, the action parametrization of dynamic parameters for such heuristic policies, (e.g., dynamical tuning $(s,S)$ values) has not yet ben extensively explored. This gap in the literature opens the door for RL methods that dynamically optimize these parameters, potentially enabling faster adoption in industry by blending RL's adaptability with the structure of traditional heuristics.

However, RL's effectiveness relies on the availability of information sharing among the entities within the supply chain where individual actors must collaborate under uncertainty and coordination constraints. In scenarios where multiple interconnected entities are involved, ensuring seamless information exchange can be challenging due to factors such as data privacy concerns, proprietary information, or communication constraints. This limitation may hinder the adoption and implementation of RL-based approaches in complex supply chain environments, highlighting the need for strategies to address information sharing challenges effectively. 
\begin{figure}
    \centering
    \includegraphics[width=0.9\linewidth]{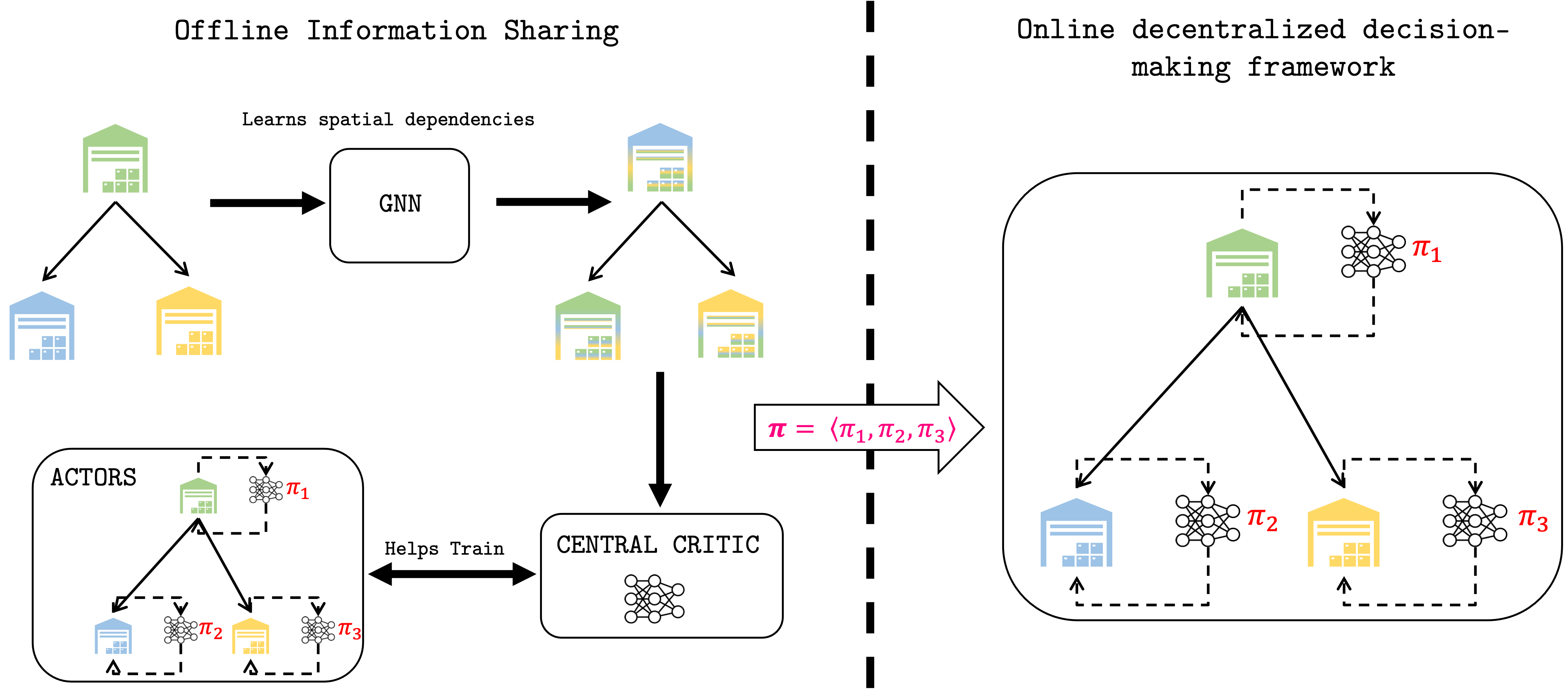}
    \caption{Multi-Agent Reinforcement Learning with Graph Neural Networks for Inventory Management – A decentralized policy learning approach where each warehouse optimizes its inventory decisions using independent neural network policies $\pi_1, \pi_2, \pi_3$ while a Graph Neural Network (GNN) captures spatial dependencies between them, enabling coordinated decision-making across the supply chain.}
    \label{fig:intro}
\end{figure}

\subsection{Multi-Agent Reinforcement Learning}
Multi-Agent Reinforcement Learning (MARL) provides a framework for addressing such challenges by enabling decentralized decision-making across multiple entities. MARL frameworks allow multiple entities within a supply chain to autonomously learn and adapt their decision-making processes while interacting with each other. By leveraging MARL, supply chain entities can collaborate and coordinate their actions more effectively, leading to improved overall performance and resilience against disruptions. Additionally, MARL provides a scalable approach to handle the complexities of large-scale supply chain systems by distributing decision-making across multiple agents. 

The application of MARL to the inventory management problem is relatively limited compared to other domains including single-agent RL. However, there is growing recognition of the potential benefits of MARL in addressing the dynamic and collaborative nature of inventory management in multi-agent environments. \citet{m1} applied Heterogeneous-Agent Proximal Policy Optimization (HAPPO) to a serial supply chain which showed overall better performance than single-agent RL. The study also concluded that information sharing between entities helps alleviate the bullwhip effect. Another study conducted by \citep{m3}, applied a Decentralized PPO framework on a single store problem with a large number of stock keeping units. This approach accelerated policy learning compared to standard MARL algorithm. Additionally, \citep{m4}  applied a PPO variation MARL framework to a multi-echelon environment, and it outperformed other base-stock policies. \citet{m2}, applied a multi-agent advantage actor critic algorithm to a multi-echelon, multi-product system but assumed a lead time of zero.  Finally, \citet{marlmousa} provided an analysis of multi-agent reinforcement learning algorithms for decentralised inventory management systems. The study showed that MAPPO outperformed other MARL methods and further highlighted MARL as a promising decentralized control solution for large-scale stochastic systems. 

However, despite its potential, MARL faces several challenges. One issue is the non-stationarity of the learning environment inherent in MARL \citep{tan1993multi, tampuu2017multiagent}. This occurs when multiple agents are learning simultaneously so the transition dynamics are not stationary. One setting commonly used to overcome this is to use a central critic during training that has access to global observations and actions \citep{nekoei2023dealing}. This is known as the Centralized Training, Decentralized Execution learning paradigm. However, while adding more information can help mitigate the non-stationarity problem, it can also lead to new issues. Naively concatenating all available information can result in information overload and inefficiencies \citep{lowe2017multi, yu2022surprising, nayak2023scalable}. This indiscriminate accumulation of data can cause policy overfitting, where agents develop strategies that perform well on the excessive training information but poorly in real-world scenarios due to the lack of generalization \citep{nayak2023scalable, hu2021policy}. Therefore, the development of novel techniques for smart information aggregation is required to avoid policy overfitting and enhance the efficacy of MARL frameworks in supply chain management.

Traditional MARL methods also lack structural awareness as they treat agent interactions as homogeneous and symmetric, failing to account for the hierarchical and graph-based structure often inherent in systems like supply chains. By failing to capture such dependencies, these methods struggle to model the nuanced coordination required in multi-entity systems. Coordination challenges are further highlighted under partial information settings, when privacy concerns or communication constraints limit the availability of data. Many MARL approaches assume full observability or reliable communication, making them ill-suited for real-world supply chains, where information exchange is often fragmented or restricted. This limitation highlights the need for methods that can operate effectively under incomplete information while leveraging structural insights to enhance coordination.

These challenges highlight the need for approaches that can leverage the inherent structure and connectivity of supply chains. Graph-based methods offer a compelling solution by explicitly modeling the hierarchical relationships and dependencies between agents. Representing supply chains as graph-structured systems—where nodes correspond to entities such as suppliers, warehouses, and retailers, and edges denote interactions like material flows or shared information—allows MARL frameworks to better capture the dynamics of multi-agent coordination.

There has been a growing interest to leverage the graph structure of a supply chain by using Graph Neural Networks (GNNs). The main idea is to leverage GNNs to learn the hidden representation of the data encoded as a graph structure \citep{sgrl1}.  This allows RL agents to efficiently adapt to changes in the problem domain. The application of GNNs to RL has been widely studied for a series of well known combinatorial optimization problems \citep{sgrl2, sgrl7} such as the vehicle routing problem \citep{sgrl1}, travel salesman problem \citep{sgrl1}and the job shop scheduling problem \citep{sgrl3, sgrl10}. GNNs have been shown to improve performance across different graph sizes and types. These works provide an initial foundation but they consider simplified supply chain instances with deterministic lead times and assume centralized information. However, real-world supply chains face inherent information sharing constraints that can arise due to technical limitations caused by incompatible systems, the inherent complex nature of supply chain structures, or even due to privacy concerns about sharing sensitive data. Therefore, there is a need to develop decentralized decision-making frameworks that not only overcome information-sharing constraints but also effectively incorporate the hierarchical and graph-based structures inherent in supply chains

\subsection{Motivation}
In this work, we propose leveraging the capabilities of GNNs to learn spatial representations of agents and their interactions. These representations are then integrated into a MARL framework to find optimal inventory policies in a multi-echelon supply chain network. Our contributions are summarized as follows:
\begin{itemize}
    \item We propose the redefinition of the action space from order replenishment to parametrize a heuristic inventory control policy where both parameters can be dynamically adjusted based on current system dynamics. This ensures early adoption of new optimization techniques due to its interpretability whilst accommodating real-world complexities. Moreover, by focusing on parameterized heuristics, we can effectively navigate the challenges that RL faces with integer decisions, such as those related to order quantities.
    \item We leverage the inherent graph structure and geometric properties of a supply chain to aid collaboration between entities in a supply chain. 
    \item We reduce dimensionality and increase scalability of the MARL-GNN framework by incorporating a global mean pooling aggregation mechanism within our algorithmic framework.
    \item We introduce Gaussian perturbations into the value function and perform a sensitivity analysis on the perturbation intensity to reduce policy overfitting and address potential distributional shift, serving as a regularization technique. 
\end{itemize}

The rest of this paper is organized as follows: Section \ref{sec:prelim} provides the background on multi-agent reinforcement learning, Section \ref{sec:method} describes our proposed decentralized decision-making framework in more detail, Section \ref{sec:r&d} discusses the experimental results obtained and finally Section \ref{sec:conc} concludes this paper and provides an outlook for future work. 

\section{Preliminaries} \label{sec:prelim}
We first provide a background into the components that contribute towards our methodology including single agent Proximal Policy Optimisation (PPO) and the multi-agent extensions of PPO. 

\subsection{Markov Decision Process (MDP)}
A Markov Decision Process (MDP) is a mathematical framework used to describe a sequential decision-making problem for a single agent interacting with an environment. An MDP is defined as a tuple $\left\langle \mathcal{S},\mathcal{A},\mathcal{T},\mathcal{R},\gamma\right\rangle$ where $\mathcal{S}$ is the set of all states, $\mathcal{A}$ is the set of actions, $\mathcal{T}: \mathcal{S} \times \mathcal{A} \times \mathcal{S}\rightarrow [0,1]$ is the state transition probability function where $\mathcal{T}(s'|s,a)$ defines the probability of transitioning to a state $s' \in \mathcal{S} \subseteq \mathbb{R}^{n_s}$ given that the agent is currently in state $s\in \mathcal{S} \subseteq \mathbb{R}^{n_s}$ and takes action $a\in \mathcal{A} \subseteq \mathbb{R}^{n_{a'}}$. For a deterministic policy $\pi$, the agent takes actions $a_t = \pi(s_t)$, while for a stochastic policy, the action is sampled from a policy $\pi$ represented by a conditional probability distribution $a_t \sim \pi(\cdot|s_t)$. $\mathcal{R}: \mathcal{S} \times \mathcal{A}\rightarrow\mathbb{R}$ is the reward function which specifies the immediate reward $r_t$ received after transitioning from state $s$ to $s'$ by taking action $a$ and $\gamma \in [0,1]$ is the discount factor. For further treatment of the subject, readers are referred to \citet{sutton2018reinforcement}.

\subsection{Single Agent Reinforcement Learning - PPO}
Proximal Policy Optimisation (PPO) is a popular first-order, on-policy\footnote{Technically, PPO employs off-policy corrections, meaning it reuses samples collected during training, but it does not explicitly use a replay buffer. PPO updates the policy network using a surrogate objective function that constrains the policy update to be within a certain proximity of the previous policy. This avoids the need to store and sample experiences} single agent reinforcement learning method. PPO is an actor-critic algorithm where a policy $\pi_{\theta}(a|s)$ and value function $V_{\phi}(s)$ are two separate neural networks parameterised by $\theta$ and $\phi$ respectively. In actor-critic algorithms like PPO, $V_{\phi}(s)$ is introduced to reduce the variance but may introduce a bias in $\pi_{\theta}$. The two variations of PPO are using a penalty function or a clipping function. The later is known to be crucial for its performance as the clipping function constraints the ratio between the new and old policy within a certain range to prevent large policy updates that may lead to instability. \\ The PPO policy loss can be defined as: 
\begin{equation}
    \mathcal{L}_{\theta} = \mathbb{E}_{(s_t, a_t) \sim \pi_{\theta_{old}}} \left[ \min \biggl(\frac{\pi_{\theta}(a_{t} | s_{t})}{\pi_{\theta_{old}}(a_{t} | s_{t})} A_{t}, \text{clip}\biggl(\frac{\pi_{\theta}(a_{t} | s_{t})}{\pi_{\theta_{old}}(a_{t} | s_{t})}, 1 - \epsilon, 1+ \epsilon\biggr)A_{t} \biggr)
\right]
\end{equation}
where $\pi_{\theta}(a_t | s_t)$ represents the probability of taking actions $a_t$ at state $s_t$ under the current policy parameterized by $\theta$, $\pi_{\theta_{old}}(a_t | s_t)$ is fixed during the policy update step and represents the probability of taking action $a_t$ at state $s_t$ under the old policy (from the previous iteration) parameterized by $\theta_\text{old}$, $a_t$ and $s_t$ are the action and state respectively taken at time step $t$, $\epsilon$ represents a hyperparameter that determines how much the new policy can deviate from the old policy and $\text{clip}(\cdot, 1 - \epsilon, 1 + \epsilon)$ is the clipping function that constraints the ratio of the new to old policy's probabilities within a certain range which prevents the policy from making overly large updates that could lead to instability. Finally, $A_{t}$ is the advantage function defined as: 
\begin{equation}
    A_{t} = R_{t} + \gamma V_{\phi}(s_{t+1}) - V_{\phi}(s_{t})
\end{equation}
where $R_{t}$ and $V_{\phi}(s_{t})$ are the reward and value function at time step $t$ respectively and $\gamma$ is the discount factor. The advantage function, $A_t$, quantifies how much better or worse the action taken at time $t$ is compared to the expected value of being in the state, effectively measuring the relative advantage of an action in improving future rewards. The general advantage estimation (GAE), used to compute the advantage, is given by: 
\begin{equation}\label{eqn: GAE}
    \hat{A}_{t}^{\textbf{GAE}} = \sum_{l=0}^{\infty}(\gamma\lambda)^{l}\delta_{t+l}
\end{equation}
where the variable $l$ represents the index of the future time step relative to the current time step $t$. This summation captures the discounted temporal difference (TD) errors, $\delta_{t+l}$, where $\delta_{t+l} = R_{t+l} + \gamma V(s_{t+l+1}) - V(s_{t+l})$ is the TD error.  GAE is commonly used in single-agent reinforcement learning as it allows for a bias-variance trade-off through its hyperparameter $\lambda$ and the summation is modulated by the term $(\gamma \lambda)^l$.  This formulation helps in accurately estimating the advantage by weighing the importance of future rewards and the associated value function estimates.

\subsection{Multi-Agent Reinforcement Learning}\label{sec:MAPPO}
Two multi-agent variants of the popular single agent PPO algorithm are: IPPO and MAPPO. 

\textbf{Independent PPO (IPPO)} IPPO is an independent learning algorithm which breaks down a problem with $n$ agents into $n$ decentralized single agent problems.  A value function, $V^i_{\phi}(s)$, and policy, $\pi^i_{\theta}$, are present for each agent in IPPO, taking local inputs. Despite showing good overall performance in certain multi-agent settings, IPPO can lead to non-stationarity in the environment. This occurs because each agent's policy is updated simultaneously which affects the state transition probability, $p(s', r | s, a^{i}, \pi)$, which becomes non-stationary. Therefore, the convergence of the Bellman Equation, shown in Equation \ref{bellman}  is not guaranteed and presents convergence problems in practice.
\begin{equation} \label{bellman}
    V^{\pi^{i}} (s) = \sum_{a} \pi^{i}(a^{i}|s) \sum_{s', r} p(s', r | s, a^{i}, \pi)(r + v_{\pi^{i}}(s')) \quad \text{for } i = 1, \ldots, n_a
\end{equation}

\textbf{Multi-agent PPO (MAPPO)}
MAPPO utilizes a centralized value function $V_{\phi}(s)$ that takes global inputs. 
The objective function can be denoted as:
\begin{align}\label{lossmappo}
\mathcal{L}(o^i, s, a^i; \mathbf{a^{-}}, \theta_k, \theta) &= \frac{1}{n_a} \sum_{i=1}^{n_a} \mathbb{E} \left[ \min \left( \frac{\pi_{\theta^i}(a^i | o^i)}{\pi_{\theta_\text{old}^i}(a^i | o^i)} A^{\pi_{\theta_\text{old}^i}}(o^i, s, \mathbf{a^{-}}), \right. \right. \nonumber \\
&\left. \left. \quad \mathrm{clip} \left( \frac{\pi_{\theta^i}(a^i | o^i)}{\pi_{\theta_\text{old}^i}(a^i | o^i)}, 1 - \epsilon, 1 + \epsilon \right) A^{\pi_{\theta_\text{old}^i}}(o^i, s, \mathbf{a^{-}}) \right) \right]
\end{align}

Where $a$ is the current agent action, $\mathbf{a^{-}}$ is the concatenated action of all agents, $s$ is the global state, $o$ is the local observation, The advantage function $A^{\pi_{\theta_\text{old}^i}}$ is computed using the GAE method in a similar manner to Equation \ref{eqn: GAE}. 

However, when GAE is applied in a multi-agent setting with a shared value function, the advantage estimated for each agent can be identical. In MAPPO, this is always true in fully cooperative environments where all agents share a common reward function and experience the same state transitions. This makes it challenging to accurately quantify and distinguish the unique contribution of each individual agent to the overall performance, even though the policy gradient considers the actions of all agents when updating the policy for each agent. This is known as implicit multi-agent credit. In contrast, in IPPO, where each agent has its own independent value function, the advantage estimates differ across agents, even in cooperative environments. This is because each agent calculates its own GAE based on its individual observations, rewards, and value function. 
\begin{equation}
    \frac{\delta \mathcal{L}}{ \delta \pi^{i}(a_{t}^{i}|s_{t})} \propto \mathbb{E} \left[ A^{\pi_{\theta_{old}}} \left(o,s, \mathbf{a^{-}}\right)\right]
\end{equation}
An additional complication is that for a large number of agents, the number of possible joint actions becomes vast, so exploring the joint action space and create enough excitation for all agents to compute the true gradient is impractical.  Therefore, learning algorithms rely on sampling techniques to estimate the gradient which may not adequately explore the joint agent space. This may lead to the problem of policy overfitting in cooperative multi-agent environments. Several studies have been conducted to reduce credit assignment issues through techniques like reward shaping \citep{zhou2020learning}, employing individual critics \citep{hernandez2019survey} and communication protocols \citep{feng2022multi}. Other methods have been developed to tackle the effects of credit assignment by reducing policy overfitting with exploration bonuses \citep{yarahmadi2023improving} or regularized policy gradients \citep{liu2021cooperative}. 

\begin{figure}
  \begin{subfigure}[b]{0.5\textwidth}
    \centering
    \includegraphics[width=\textwidth]{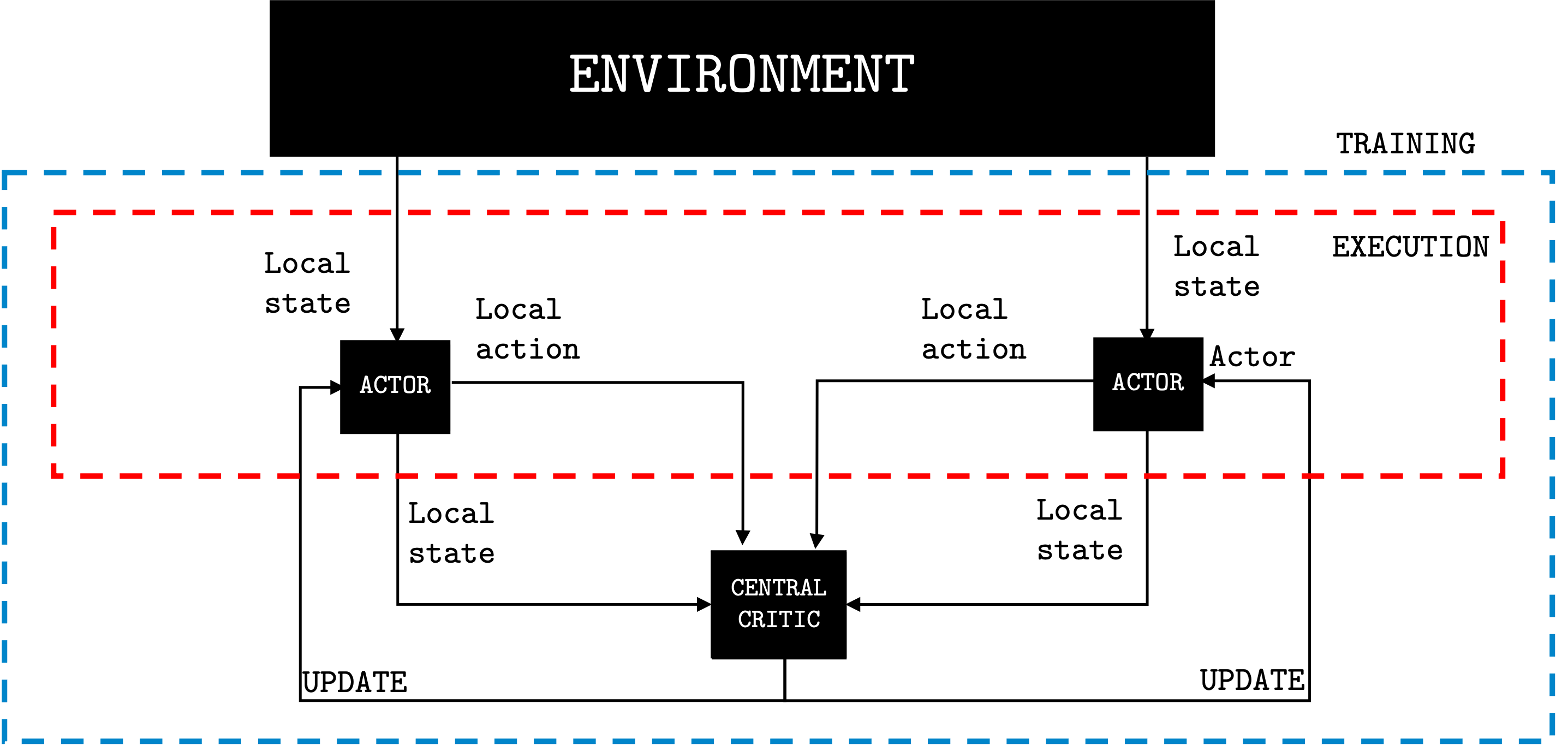}
    \caption{Abstract Representation (Black-Box)}
    \label{fig:ctde}
  \end{subfigure}
  \hfill
  \begin{subfigure}[b]{0.5\textwidth}
    \centering
    \includegraphics[width=\textwidth]{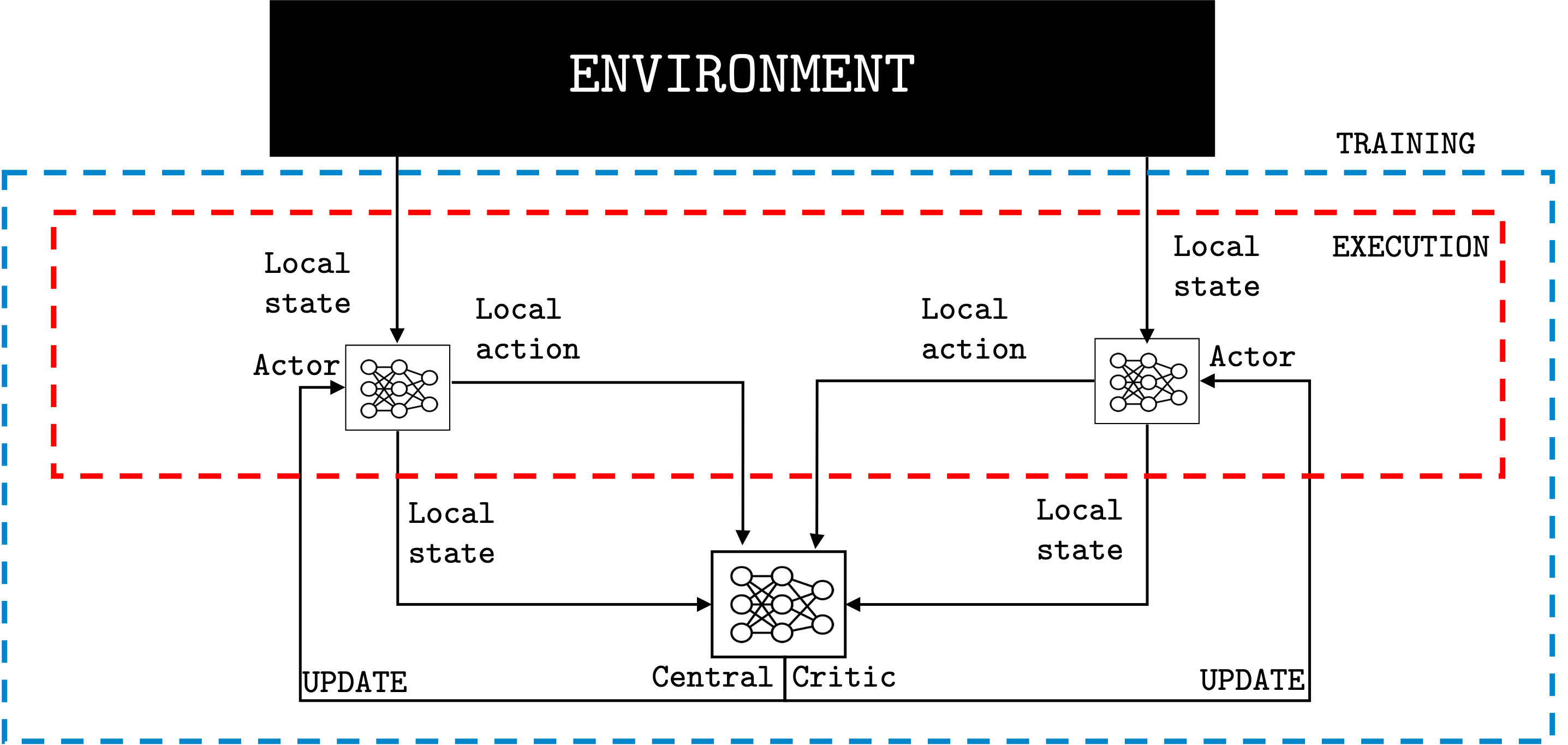}
    \caption{Parametrized by Neural Networks}
    \label{fig:ctdeb}
  \end{subfigure}
  \caption{Schematic showcasing the Centralized Training Decentralized Execution framework, demonstrating how agents undergo collaborative training in a centralized manner (blue dashed line) while executing actions independently in their respective environments (red dashed line). }  
  \label{fig:overall_ctde}
\end{figure}

\textbf{Centralized Training Decentralized Execution (CTDE).} CTDE is a framework used in cooperative MARL to overcome some of the shortcomings mentioned above. In this setting, agents take global state information in a centralized manner to help train policies that can execute on a decentralized manner at execution with local inputs only as shown in Figure \ref{fig:overall_ctde}. Our paper focuses on a cooperative MARL setting where agents only share a common reward function. It is widely known that sharing information between agents helps stabilize learning and deals with the non-stationarity problem inherent in multi-agent problems. Despite the increased performance, sharing information by naïvely concatenating local information leads to the curse of dimensionality as the global state increases with the number of agents. 

\section{Methodology} \label{sec:method}
In this section, we outline our approach for a graph-based multi-agent PPO algorithm. First, we define the inventory management problem, outlining its key components and objectives. Next, we discuss the graph representation of our problem. Finally, we describe our proposed framework with  the integration of  graph neural networks (GNNs) and illustrate how noise injection is utilized as a regularizer within the value function in our approach. 

\subsection{Problem Statement}

To find the optimal inventory policy in the inventory management problem, we propose a mathematical formulation of the supply chain dynamics as an optimization problem, characterized by each time period $t$, over a fixed horizon of $T$ time periods. The variables are defined in Table \ref{tab:inventory_management}. 

\begin{table}[h]
    \centering
    \caption{Variables and Parameters for the Inventory Management Problem}
    \begin{tabular}{@{}ll@{}}
        \toprule
        \textbf{Symbol} & \textbf{Description} \\ \midrule
        \(i\) & Node \(i \in N\) where \(N\) is the total number of nodes \\
        \( g \in \mathbb{R}^{n_{i_d}} \) & Amount of goods shipped from node \( i \) to downstream nodes \\ 
        \( o_r \in \mathbb{R} \) & Replenishment order \\ 
        \( d \in \mathbb{R} \) & Demand from downstream nodes \\ 
        \( v \in \mathbb{R} \) & On-hand inventory \\ 
        \( b \in \mathbb{R} \) & Backlog \\ 
        \( q \in \mathbb{R} \) & Acquisition or incoming goods \\ 
        \( v_0 \in \mathbb{R} \) & On-hand inventory at the start of each period \\ 
        \( b_0 \in \mathbb{R} \) & Backlog at the start of each period \\ 
        \( P \in \mathbb{R}^{n_N} \) & Price of goods sold \\ 
        \( C \in \mathbb{R}^{n_N} \) & Order replenishment costs \\ 
        \( V \in \mathbb{R}^{n_N} \) & Storage costs \\ 
        \( B \in \mathbb{R}^{n_N} \) & Backlog costs \\ 
        \( V_{\max} \in \mathbb{R}^{n_N} \) & Maximum limits on node storage \\ 
        \( O_{r,\max} \in \mathbb{R}^{n_N} \) & Maximum limits on replenishment order quantities \\ 
        \( i_u \in \mathbb{R}\) & Upstream node of \( i \) \\ 
        \( i_d \in \mathbb{R}\) & Downstream node of \( i \) \\
        \( b^i = \sum_{j \in \mathcal{D}_i} b^{i_d} \) & Total backlog of node \( i \) from downstream nodes  \(j\)\\ 
        \( g^i = \sum_{j \in \mathcal{D}_i} g^{i_d} \) & Total shipment of node \( i \) to downstream nodes \(j\)\\ 
        \( \mathcal{D}_i \) & Set of direct downstream nodes \(j\) of node \( i \) where \(j \in \mathcal{D}_i\)\\ 
        \( \mathcal{C} \) & Set of nodes with customer demand \\ 
        \( c \) & Customer demand \\ 
        \bottomrule
    \end{tabular}
    \label{tab:inventory_management}
\end{table}

\begin{figure}
    \centering
    \includegraphics[width = 0.5 \textwidth]{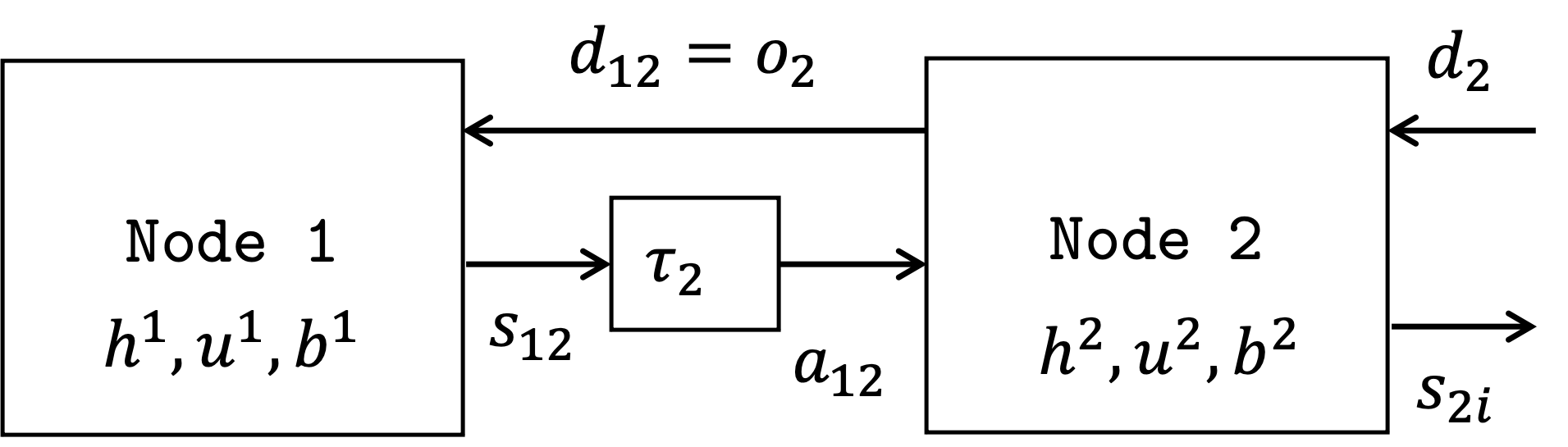}
    \caption{Schematic showing the inventory flow between two nodes in an inventory management system}
    \label{fig:invflows_}
\end{figure}

Equation (\ref{eqn: optimisation}) is the objective function that maximizes total profit across the supply chain system. The system is treated as a collaborative framework, where all agents in the network share a common objective function, reward $\mathcal{R}_t$. The decision variables in this optimization include $o_{r}^{i}[t]$ which represents the order quantity at node $i$ at time $t$. Other decision-related quantities, like inventory and backlog, are derived from this order quantity based on the system's dynamics, as captured by the relevant constraints. Equation (\ref{eqn: inventory}) and (\ref{eqn: backlog}) show how the inventory and backlog are updated over time. Equations (\ref{eqn: sales constraint 2}) and (\ref{eqn: sales constraint 1}) restrict the quantity of goods a node can ship downstream, ensuring it does not exceed the on-hand inventory or the downstream demand and backlog. Equation (\ref{eqn: acquisition}) and (\ref{eqn: factory}) capture the lead time of a shipment, indicating that goods shipped to node \( i \) will take \( \tau^i \) periods to reach the downstream stage. The amount of inventory hold or ordered is also constrained to a maximum value through Equations (\ref{eqn: limits}).  The interaction of the different flows between two nodes can also be seen in Figure \ref{fig:invflows_}. 

The overarching goal is to maximize the net profit generated across all nodes and all time periods. We can formalize the optimization problem as follows:
\begin{align}
\begin{split}\label{eqn: optimisation}\scriptscriptstyle
&\max \sum_{i=1}^N \sum_{t=1}^T P^i g^i[t] - C^i o_{r}^i[t] - V^i v^i[t] - B^i b^i[t] \,, 
\end{split}\\
\begin{split}
&\text{subject to:} \notag 
\end{split}\\
\begin{split}
\label{eqn: inventory}
&v^i[t] = v_0^i[t] - g^i[t] + q^i[t]\,, \quad \forall i, \forall t, 
\end{split}\\
\begin{split}
\label{eqn: backlog}
&b^{i_d}[t] = b_0^{i_d}[t] - g^{i_d}[t] +  d^{i_d}[t], \quad \forall i, \forall d \in \mathcal{D}_i\,,
\end{split}\\
\begin{split}
\label{eqn: sales constraint 2}
&g^{i_d}[t] \leq b_0^{i_d}[t] + d^{i_d}[t]\,, \quad \forall i, \forall t, \forall d \in \mathcal{D}_i 
\end{split}\\
\begin{split}
\label{eqn: sales constraint 1}
&g^i[t] \leq v_0^i[t] + q^i[t],  \quad \forall i, \forall t,    
\end{split}\\
\begin{split}
\label{eqn: acquisition}
&q^i[t] = g^{i_u}[t-\tau^i], \quad \forall i \neq 1\,, t\geq \tau^i  
\end{split}\\
\begin{split}
\label{eqn: factory}
&q^1[t] = o_r^1[t-\tau^1], \quad t\geq \tau^1
\end{split}\\
\begin{split}
\label{eqn: order demand equality}
&d^{i_d} = o_{r}^d,  \quad \forall i, \forall d \in \mathcal{D}_i\,,  
\end{split}\\
\begin{split}
&\text{with} \notag 
\end{split}\\
\begin{split}
\label{eqn: customer demand}
& d^i[t] = c^i[t], \quad \forall i \in \mathcal{C}, \forall t   
\end{split}\\
\begin{split}
\label{eqn: limits}
&o_{r}^i[t] \leq O_{r,\max}^i\,, v^i[t] \leq V_{\max}^i\,, \quad \forall i, \forall t,  
\end{split}
\end{align}
\normalsize

In our model, we take into account exogenous forms of uncertainty on the demand and lead time , modeled as a Poisson random variable with parameter $\lambda_\text{d}$ and $\lambda_\text{l}$ respectively. While inventory and backlog are treated deterministically to simplify the analysis and focus on long-term system performance, the framework indirectly incorporates the effects of uncertainty by modeling demand and lead time as stochastic variables - two key sources of uncertainty. This approach strikes a balance between computational efficiency and the need to capture essential uncertainties. Therefore, the probability of $k$ demands in a time period $t$ is given by:

\begin{equation}
    P(D=k)= \frac{(\lambda t)^{k}\exp^{-kt}}{k!}
\end{equation}

Moreover, we model the inventory control problem as a sequential decision-making problem where each node is modeled as a separate agent. The goal is to determine the optimal order quantity for each supply chain entity at each time step, ensuring coordination across agents while maximizing overall profit. In the context of multi-agent reinforcement learning, we model the supply chain as cooperative which means agents coordinate towards a common goal, receiving a shared reward. This sequential decision making problem is modelled as an extension of the MDP framework introduced in Section \ref{sec:prelim}, adapted to a multi-agent and partially observable setting as a Decentralized Partially Observable Markov Decision Process (Dec-POMDP), which can be defined as a tuple $\left\langle \mathcal{S},\mathbf{A},\mathcal{T},\mathcal{R},\Omega,\mathcal{O},\gamma\right\rangle.$ $\mathcal{S}$ is the set of all valid states representing the joint state space shared by all agents, $\mathbf{A} \coloneqq \mathbf{A}^{1} \times \dots \times \mathbf{A}^{n_a}$ is the joint action space where $\mathbf{A}^{i} $ is the set of actions available for each agent $i$, $\Omega \coloneqq \Omega^{1} \times \dots \times \Omega^{n_a}$ is the joint observation space and $n_a$ denotes the number of agents. At each time step, each agent $i$ executes action  ${a}^{i} \in\mathcal{A}^{i} \subseteq \mathbb{R}^{n_{a'}}$ with a joint action $\textbf{a} = \left\langle a^1,\dots,a^{n_a}\right\rangle$ and transitions from state $s\ \in\mathcal{S} \subseteq \mathbb{R}^{n_s}$ to $s^\prime \in\mathcal{S} \subseteq \mathbb{R}^{n_s}$ with state transition probability $P(s^\prime|s, \textbf{a}) = \mathcal{T}(s,\textbf{a},s^\prime)$. Each agent $i$ receives observation $o^i \in\Omega^i \subseteq \mathbb{R}^{n_{o}}$ determined by $\mathcal{O}(s^\prime, i)$ which maps the new state $s^\prime \in \mathcal{S}$ to an observation $o^i \in \Omega^i$ for each agent $i$. In other words, the observation function $\mathcal{O} : \mathcal{S} \times \{ 1, \dots, n_{a} \} \rightarrow \Omega$ provides each agent $i$ with a local observation $o^i$ based on the next state $s^{\prime,i}$. The joint observation can be defined as $\textbf{o} = \left\langle o^1,\cdots,o^{n_a}\right\rangle$ and each agent shares the same reward function $\mathcal{R}(s,\textbf{a})\in\mathbb{R}$. Each agent has policy $\pi^i$ and the joint policy is denoted as $\mathbf{\pi}=\ \left\langle\pi^1,\cdots,\ \pi^{n_a}\right\rangle$. The optimal joint policy is found through maximizing the joint expected reward $\mathbb{E} \left[\sum_{t=0}^{t=T}\gamma^tr_t\right]$ where $r_t\ =\ \mathcal{R}\left(s_t,\ a_t\right)\ \in\mathbb{R}$ at each time step $t$ where $T$ is the time horizon and the discount factor $\gamma\in\left[0,1\right]$.

Each inventory control agent within the system is characterized by the following attributes: \\
\textbf{State Space.} In the context of our inventory control system, the observation set for each agent $i$ is $o^{i} = [v, b, p, d^{-1}, \dots, d^{-M}, o_{r}^{-1}, \dots, o_{r}^{-M}]$ where $o^i\ \in\Omega^i$. In the observation set, a new variable is introduced, $p$, which is the pipeline inventory equal to the sum of order replenishment that has not yet arrived at the node from other upstream nodes. 

To mitigate the problem of partial observability, we include demand history and order history up to $M$ time-steps in the past where $M$ is a hyperparameter. While Recurrent Neural Networks (RNNs) could handle sequential data and temporal dependencies, we chose not to use them here for simplicity and practicality in training. Our decision to include a fixed window of past observations instead allows for a simpler implementation while still capturing relevant historical context. It is important to note that this inclusion of historical data introduces a violation of the Markov property, which states that the future state of the system only depends on the current state and not on the sequence of events that preceded it. However, real-world decision-making processes rarely exhibit perfect Markovian behavior.  Therefore, including historical data is a well-established approach to augment the observation space in partially observable environments \citep{liu2022partially, marlmousa, uehara2022provably}. \\
\textbf{Action Space.} The action space is traditionally modeled as the order replenishment quantity, $o_r$.  While the actual order replenishment quantities in our environment are discrete, we model the action space as continuous in  $\left[-1, 1\right]$ for scalability to a wider range of possible order sizes. In this paper we parameterize a heuristic inventory policy, specifically an $(s,S)$ policy, using a neural network policy. The $(s,S)$  policy is defined by two key parameters: the reorder point $s$, which triggers a replenishment order when the inventory reaches a specific level, and the order-up-to level $S$, which is the target inventory level after replenishment. The replenishment quantity, $o_r$, is dynamically determined as the difference between the order-up-to level and the current inventory level. Unlike traditional implementations that rely on fixed values for $s$ and $S$, we extend this approach by parameterizing $s$ and $S$ as stochastic variables drawn from Gaussian distributions, allowing for dynamic adaptation to demand and lead-time uncertainties.

For each agent $i$, at each time step $t$, the policy outputs the mean and standard deviation for both the reorder point $s$ and order-up-to level $S$. The policy is defined as follows:
\begin{align}
    \pi_{i}(o^{i}) = (\mu_{s_{\text{inv},t}^i}, \sigma_{s_{\text{inv},t}^i}, \mu_{S_{\text{ord},t}^i}, \sigma_{S_{\text{order},t}^i})
\end{align}
where the reorder point $s_{\text{inv},t}^{i}$ and the order-up-to level $S_{\text{ord},t}^{i}$ are sampled from a Gaussian distribution:
\begin{align}
    s_{\text{inv},t}^i \sim \mathcal{N}(\mu_{s_{\text{inv},t}^i}, \sigma_{s_{\text{inv},t}^i}^2) \\
    S_{\text{ord},t}^i \sim \mathcal{N}(\mu_{S_{\text{ord},t}^i}, \sigma_{S_{\text{ord},t}^i}^2)
\end{align}

A min-max post processing step is then used to scale the values to a suitable range denoted by a subscript $s$, leading to $(s_{t_\text{inv,s}}^{i}, S_{t_\text{ord,s}}^{i})$ where $s_{t_\text{inv,s}}^{i}$ is the reorder point and $S_{t_\text{ord,s}}^{i}$ is the order-up-to-level. The min-max scaling step is defined as:
\begin{align}
    f(s, s_{\text{min}}, s_{\text{max}}) &= \frac{s+1}{2} \times (s_{\text{max}} - s_{\text{min}})+s_{\text{min}} \\
    s_{t_\text{inv,s}} &= f(s_{\text{inv},t}, s_{\text{inv, max}}, s_{\text{inv, min}}) \\
    S_{t_\text{ord,s}} &= f(S_{\text{ord},t},S_{\text{ord, max}}, S_{\text{ord, min}})
\end{align}
where $s_{\text{inv,min}}$, $s_{\text{inv,max}}$, $S_{\text{ord,min}}$ and $S_{\text{ord,max}}$ represent the lower and upper bounds for $s$ and $S$ respectively.

\begin{figure}
    \centering
    \includegraphics[width=\textwidth]{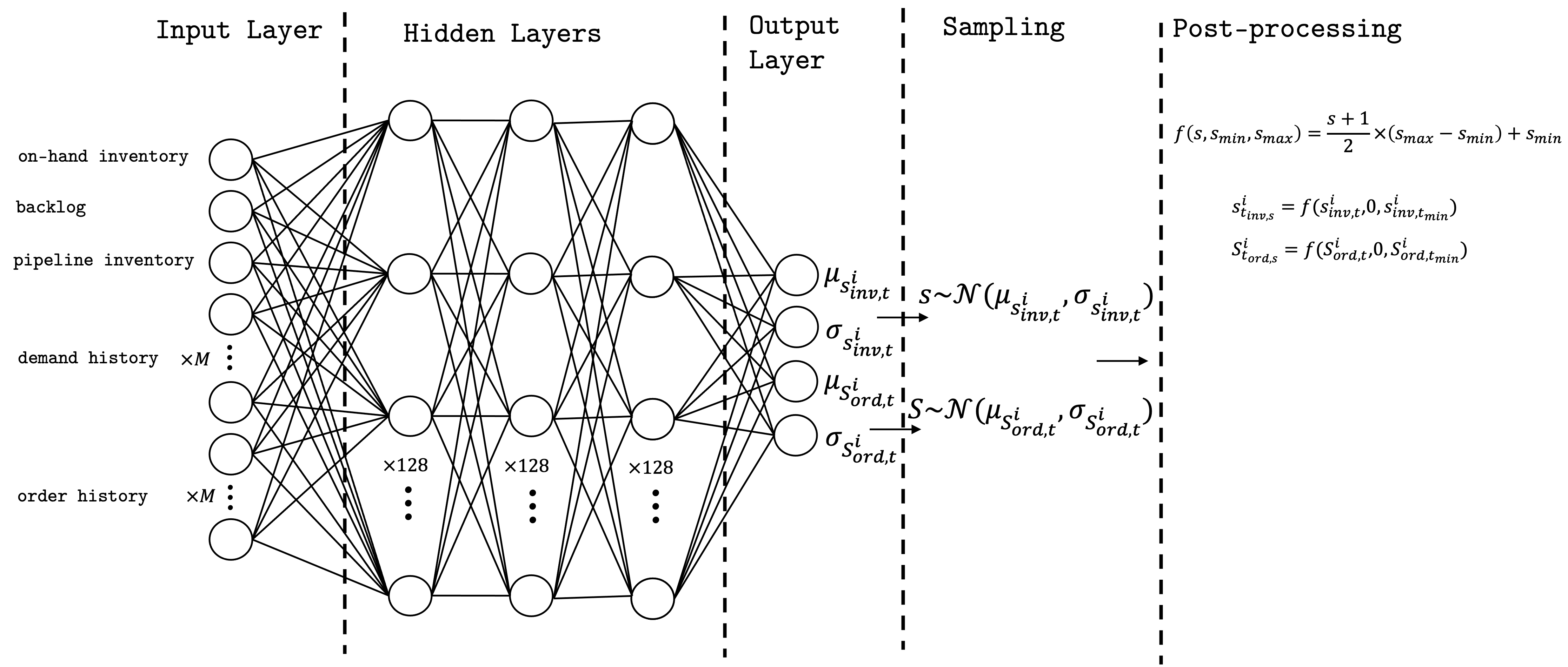}
    \caption{Neural network architecture for actor, illustrating sampling from a Gaussian distribution, followed by a post-processing step, leveraging an inventory heuristic policy to generate actions in a continuous action space.}
    \label{fig:nn_architecture}
\end{figure}

Once scaled, the reorder point  $s_{t_\text{inv,s}}^{i}$ and the order-up-to level $S_{t_\text{ord,s}}^{i}$ are used to determine the order quantity. When the inventory reaches a level of $s_{t_\text{inv,s}}^{i}$, an inventory order is placed where $o_{r,t}^{i} = S_{t_\text{ord,s}}^{i} - v_{t}^{i}$ , rounded to the nearest integer. The neural network architecture for each actor, including the post-processing step that results in the order replenishment quantity, is illustrated in Figure \ref{fig:nn_architecture}. 

Moreover, in inventory management, optimal order quantity policies such as $(s_{t_\text{inv,s}}^{i}, S_{t_\text{ord,s}}^{i})$ are often characterized by discrete functions as they show abrupt changes in order quantity at specific inventory levels \citep{dehaybe2024deep}. This poses a challenge for neural networks, which are inherently continuous function approximators as directly approximating such discrete policies can lead to instability and poor performance in neural network-based models. To address this, in this paper, the actor network outputs a normal distribution (Gaussian) over the $(s_{t_\text{inv}}^{i}, S_{t_\text{ord}}^{i})$ parameters. A key note is that this approach does not limit the network's ability to learn the optimal policy as it simply expresses the policy in a way that may be easier to learn. Moreover, unlike traditional heuristic methods where the policy is not conditioned on the state of the system, an RL approach means the heuristic $(s_{t_\text{inv,s}}^{i}, S_{t_\text{ord,s}}^{i})$ policy is conditioned on the state of the system such as on-hand, pipeline inventory and backlog. This enables the agent to dynamically adjust the reorder point and order-up-to level based on the current inventory situation, potentially leading to optimal and flexible decision-making.

\subsection{Graph Representation of Supply Chain Systems}\label{sec:graphs}
Graph-based systems are commonly used in various domains to model complex networks of interactions. For example, transportation networks, social networks, recommendation systems, and communication networks all rely on graph representations to model the connections and relationships between different entities. Similarly, in supply chain management, viewing the system as a graph allows us to better understand and optimize the flow of goods, information, and resources across interconnected entities.

A supply chain can naturally be represented as a graph $G = (W,E)$ where the entities within the supply chain are represented as nodes $i \in W$, and the relationships or interactions between these entities are represented as edges $E$. This graph representation provides a framework for modeling the complex interactions and dependencies that characterize inventory management systems. 

To effectively apply this graph-based approach, it is essential to translate a real-world supply chain into the corresponding graph representation. The key components of this translation are outlined below:\\
\textbf{Nodes (Entities)} Each node $i \in W$ corresponds to an agent responsible for managing inventory at that entity's location where $W$ represents different entities in the supply chain.. \\
\textbf{Nodes Features} Each node $i$ has a node feature $x^{i} \in X$ which represents the relevant observations for the agent located at that node. These features capture the critical variables needed for decision-making at each entity. In the context of inventory management, these features include the current inventory level, backlog, pipeline inventory, historical demand and order history. \\
\textbf{Edges} The edges $E$ in the graph represent the relationships of interactions between the different entities. These relationships include direct transportation links, supply routes, or communication channels between entities. In this work, an edge exists between two nodes if they are connected and if the flow of goods or orders is permitted between them. \\
\textbf{Neighborhood} The neighborhood of a node $i$, denoted by $M(i)$ is defined as the set of neighboring nodes $j$ connected to node $i$ via edges in $E$.  Mathematically, this can be represented as $M(i) = \{j | (i,j)\in E \}$. These neighboring nodes represent other entities that directly interact with the current node in the supply chain. For each node $i$, the features of its neighbors $j$ are represented as $x_{j}^{i}$, capturing the relevant observations of the neighboring agents and providing the context for decision-making at node $i$.

This graph-based representation provides a flexible framework that can be applied to a wide variety of supply chain settings. By modeling the supply chain as a graph, we can capture the complex interactions between entities, allowing us to further improve collaboration when coordinating decisions across different agents. 

\subsection{\textbf{G}raph \textbf{C}onvolutional \textbf{N}etworks (GCNs) combined with \textbf{M}ulti-\textbf{A}gent \textbf{P}roximal \textbf{P}olicy \textbf{O}ptimization (MAPPO) and a \textbf{P}ooling strategy (P-GCN-MAPPO) }\label{sec:gcn_method}

In environments where the graph structure can be leveraged, Graph Neural Networks (GNNs) are commonly integrated into RL frameworks. GNNs were developed to efficiently leverage the structure and properties of graphs. GNNs operate on graph-structured data and are able to capture complex relationships and dependencies inherent in graphs. In this work, we use \textbf{G}raph \textbf{C}onvolutional \textbf{N}etworks (GCNs) combined with \textbf{M}ulti-Agent \textbf{P}roximal \textbf{P}olicy \textbf{O}ptimization (MAPPO) and a \textbf{P}ooling strategy, hence P-GCN-MAPPO. In Section \ref{sec:r&d} we conduct computational experiments to analyze the different components of this methodology in an effort to distill their contribution to the overarching framework. GCNs update the representation of a node by aggregating and transforming the features of its neighbouring nodes and itself. This allows the model to capture and propagate local information, effectively learning patterns and dependencies from the graph structure of a supply chain. 

Traditional model-free methods rely solely on agent-environment interactions, whereas our GNN-based framework incorporates geometric information from the supply chain’s topology to enhance learning \citep{almasan2022deep, zhou2020graph}. While standard model-free approaches can struggle with capturing complex dependencies, Graph Neural Networks (GNNs) leverage structural relationships to learn richer representations, even in the absence of a perfect system model \citep{yang2022graph}. This ability makes GNNs well-suited for supply chain problems, which are characterized by inherent uncertainties and complex interactions. Moreover, GNNs are robust to noisy and imperfect data, effectively capturing the underlying relationships in the presence of missing or uncertain information \citep{verma2021graphmix, jin2021adversarial}. This ability to capture complex relationships is particularly beneficial for supply chain problems as understanding the intricate relationships between various entities can significantly optimize decision-making. 

\begin{figure}
    \centering
    \includegraphics[width = \textwidth]{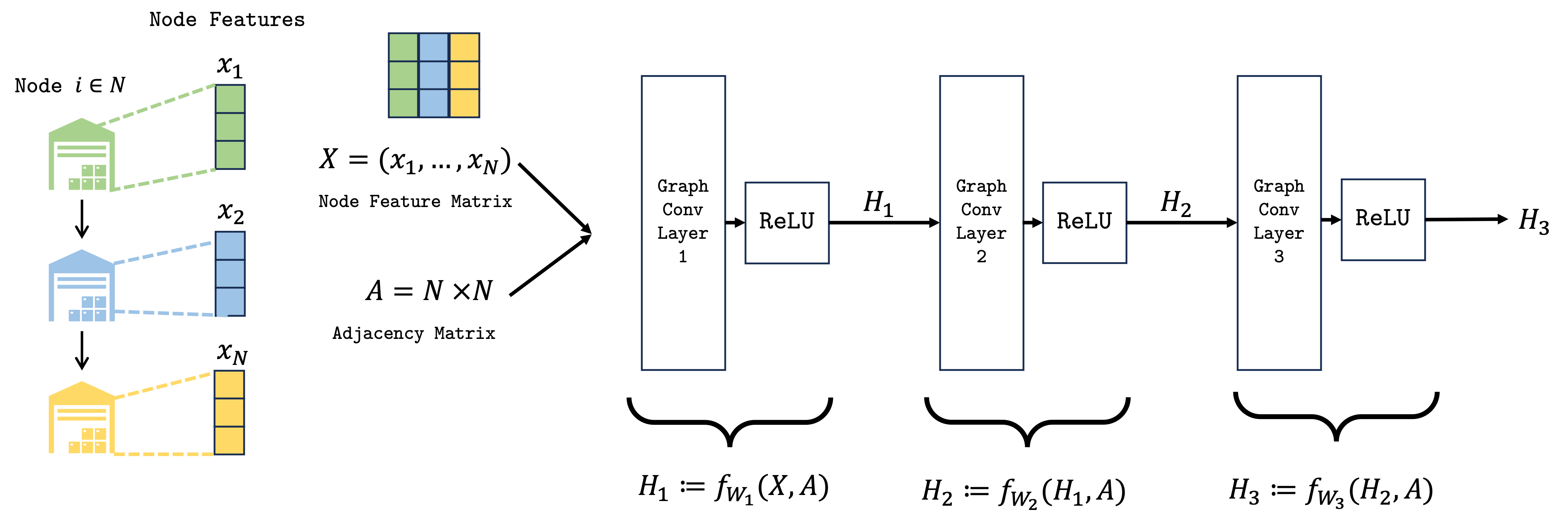}
    \caption{Illustration of the Graph Convolutional Network (GCN) architecture, representing the Graph Module of our framework. The GCN takes the adjacency matrix and node feature matrix as input, processes them through three convolutional layers, each with ReLU activation, and outputs the embedded node matrix representing the learned representations of the nodes.}
    \label{fig:gcn}
\end{figure}

The adjacency matrix, $A \in \mathbb{R}^{N \times N}$, is a $N \times N$ matrix which is used to express the directed graph topology where $N$ is the number of nodes (or vertices) in the graph. In this matrix, $A_{ij} = 1$ indicates that there is an edge between node $i$ and node $j$. Each node $i$ is associated with a node feature $x_{i}$ as described in Section \ref{sec:graphs}, which encapsulates information specific to that node where $x_i \in \mathbb{R}^D$ where $D$ represents the dimensionality of the features vector. These individual node features collectively form the node feature matrix, $X \in \mathbb{R}^{N \times D}$. At each time step, the node feature matrix, $X$, captures the evolving state of the graph. Both $A$ and $X$ are fed into a graph convolution layer, allowing the model to capture relational information between nodes. The function $f(X,A)$ represents the graph convolution operator which aggregates and transforms the node feature matrix $X$ based on the connectivity defined by $A$. This function captures the relational information between nodes and can be defined as:
\begin{equation}
    f(X,A) \coloneqq \sigma_{\text{NN}} (D^{-\frac{1}{2}}(A + I)D^{-\frac{1}{2}}XW)
\end{equation}
where $A \in \mathbb{R}^{N \times N}$ is the adjacency matrix, $I \in \mathbb{R}^{N \times N}$ is the identity matrix, $D \in \mathbb{R}^{N \times N}$ is the degree matrix of $A+I$, $X \in \mathbb{R}^{N \times D}$ is the node feature matrix, $W \in \mathbb{R}^{D \times W'}$ is the layer's weights where $W'$ is the number of output features and $\sigma_{\text{NN}}(\cdot)$ is the activation function (e.g. ReLU). This results in an embedded vector for each node, $h_i \in \mathbb{R}^{W'}$. In this work, three convolution layers are used where the embedded vector at each node, $h_i$, is the input of the next layer. This is described mathematically as: 
\begin{align}
    H_{1} &\coloneqq f_{W_{1}}(X,A) \\
    H_{2} &\coloneqq f_{W_{2}}(H_{1},A) \\
    H_{3} &\coloneqq f_{W_{3}}(H_{2},A)
\end{align}
where $H_{1} \in \mathbb{R}^{N \times W_{1}'}, H_{2}\in \mathbb{R}^{N \times W_{2}'}, H_{3}\in \mathbb{R}^{N \times W_{3}'}$ are the embedded node matrices at layers 1,2 and 3 and $W_{1}, W_{2}, W_{3}$ are the weight matrices that parameterize each layer. The terms $W_{1}', W_{2}', W_{3}' \in \mathbb{Z}^{+} 
$ are defined as positive integers representing the number of output features for each layer respectively. These values are hyperparameters that determine the dimensionality of the embedded vectors produced by each layer. The Graph Convolutional Network (GCN), which is our graph module in the framework is used to learn hidden features that capture the structural information of the graph is shown in Figure \ref{fig:gcn}.

The resulting latent representations are then passed to our centralized value function $V_{\phi}(s)$, a fully connected network parameterized by $\phi$. This function assesses the value of actions, aiding in variance reduction of the policy $\pi_{\theta}$, another fully connected network parameterized by $\theta$, which maps states to probability density functions of actions. To train the policies, we employ Multi-Agent Proximal Policy Optimization (MAPPO). Mini-batches are sampled, and the objective is to minimize the loss function for each batch shown in Equation \ref{lossmappo} which is used to update the parameters, $\theta$ and $\phi$.  

A challenge of this method is that as the number of agents increases, the dimensionality of the node feature changes. To address this, we propose integrating global mean pooling within our framework, ensuring dimensionality remains constant as the number of agents increases. 

Pooling operators in graphs were inspired by pooling methods in Convolutional Neural Networks (CNNs). Instead of simply concatenating all the latent hidden representations and feeding them directly into our central value function, we employ the global mean pool operator. The global mean pool operator aggregates information across all nodes in a graph by taking the average of the node features, providing a global representation of the graph without bias toward any specific node or agent. This approach ensures that the representation remains consistent, maintaining dimensionality across the fixed number of agents in the supply chain. The GNN outputs hidden feature matrix, $H \in \mathbb{R}^{N \times W_3'}$ ,  can be denoted as $H = \{h_1, h_2, \dots, h_{N}\}$ where $N$ is the number of nodes and $h_{i}$ is the GNN output feature vector for node $i$. The global mean pooling operator computes the mean of each feature across all nodes. 
\begin{equation}
    \text{Global Mean Pool}(H) = \frac{1}{N} \sum_{i=1}^{N}h_{i}
\end{equation}

Alternatives to global mean pooling, including global max pooling, which could ignore contributions from ``less active'' nodes, potentially overlooking useful information from nodes that play a less prominent role in the system. Attention-based pooling dynamically weighs the importance of each node, which allows for a more flexible representation that can adapt to changing system dynamics. However, this introduces additional computational complexity and requires learning more parameters, increasing the risk of overfitting and making the model harder to optimize. The reader is referred to \citet{grattarola2022understanding} and \citet{liu2022graph} for further reading on different pooling strategies.

This effectively reduces the input dimensions to our critic from $N \times W_3'$ to $1 \times W_3'$. While this dimensionality reduction improves computational efficiency, there is a potential drawback: it may result in the loss of specific spatial information. However, since each individual actor still leverages local information for each node, the loss is confined to the critic, whose role is mainly to provide global information. Moreover, the critic might not require the full level of spatial detail to assess the overall value of a state. The high-level features extracted through the global mean pooling mechanism might suffice. This idea has been explored in several studies such as \citet{fujimoto2018addressing} which highlights the critic's focus on a good enough value approximation for policy improvement, suggesting it may not need full state representation, and \citet{lyu2023centralized} which proved that centralized critics may not be beneficial; particularly state-based critics can introduced unexpected bias. This aligns with the idea that the critic might not need a perfect representation but an accurate enough estimate for policy guidance. Finally, global mean pooling mechanisms can mitigate the risk of overfitting, a common problem in RL where the model performs well on training data but poorly on unseen data. By reducing the number of features, the model focuses on the most relevant aspects, which can lead to better generalization. 

\begin{figure}
    \centering
    \includegraphics[width = \textwidth]{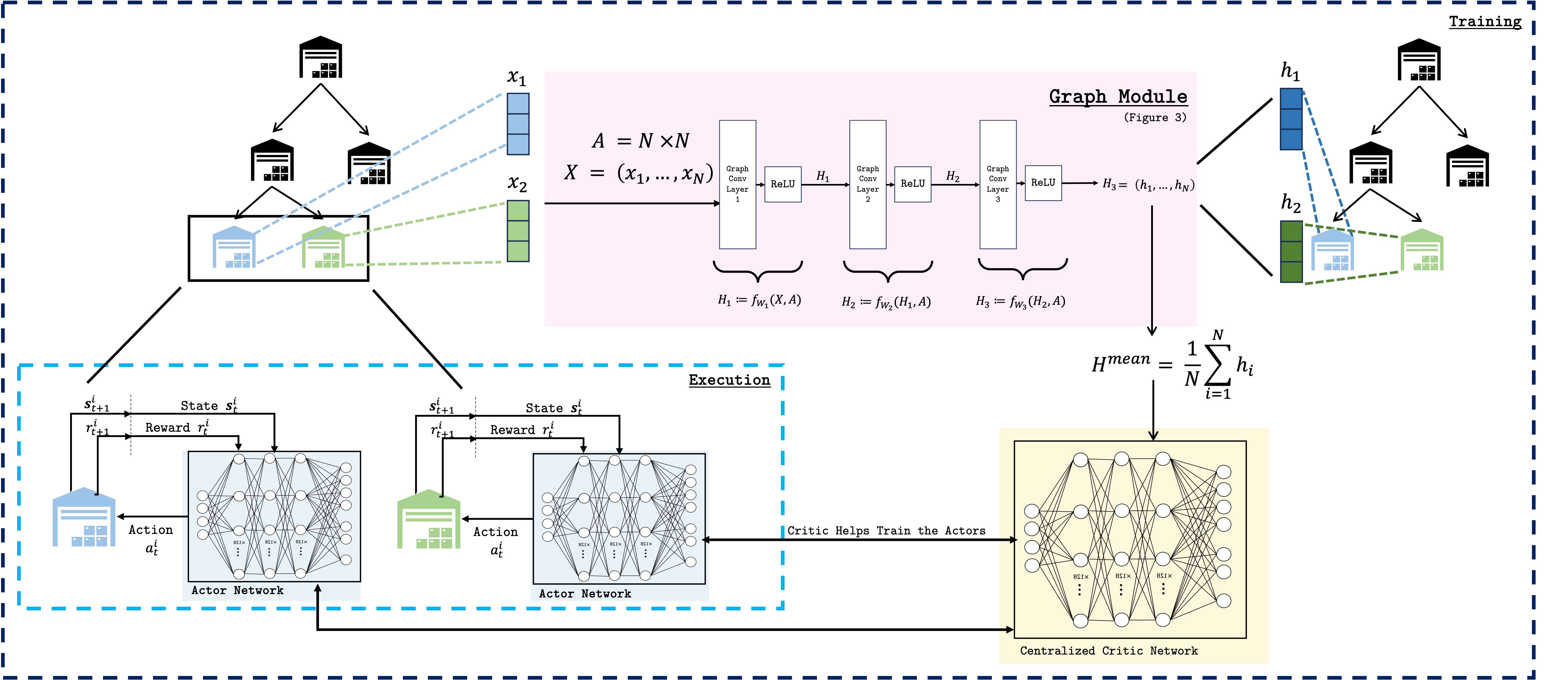}
    \caption{Methodology diagram outlining the P-GCN-MAPPO framework, illustrating the key components and processes involved.}
    \label{fig:GPMAPPO}
\end{figure}

Figure \ref{fig:GPMAPPO} illustrates the framework which integrates Graph Convolutional Networks (GCNs) with Multi-Agent Proximal Policy Optimization (MAPPO) to enhance coordination among agents in a structured environment. The system is composed of three key components: the graph-based representation, the centralized critic, and the actors. 

The environment is modeled as a graph, where nodes represent individual agents, and edges capture interactions between them as described in Section \ref{sec:graphs}. Each agent's local state information is encoded in a node feature matrix, while the adjacency matrix defines connectivity based on the edges (i.e., based on what supply chain nodes are connected). A Graph Convolutional Network (GCN) processes this structured input to extract informative representations by aggregating features from neighboring nodes. 

The output of the GCN undergoes a global mean pooling operation, producing a single aggregated representation of the system's state. This representation serves as input to the centralized critic, which estimates a shared value function to guide policy optimization. By leveraging this centralized value estimation, the approach facilitates effective coordination while maintaining decentralized execution.

Each agent is equipped with an actor network that receives local state information and outputs an action. These policies are trained using MAPPO, where the critic provides a learning signal to optimize the policies while maintaining stability in multi-agent settings as described in Section \ref{sec:MAPPO}. During training, both the actor and critic networks are updated, with the critic leveraging the pooled global state to inform policy learning. However, during execution, only the decentralized actor networks are utilized, ensuring scalability and real-time decision-making without reliance on a central entity.

This architecture effectively combines graph-based feature extraction with multi-agent reinforcement learning, enabling agents to capture relational dependencies while optimizing their policies in a structured and scalable manner.

\subsection{\textbf{R}egularized \textbf{G}raph \textbf{C}onvolutional \textbf{N}etworks (GCNs) combined with \textbf{M}ulti-\textbf{A}gent \textbf{P}roximal \textbf{P}olicy \textbf{O}ptimization (MAPPO) and a \textbf{P}ooling strategy (Reg-P-GCN-MAPPO)}\label{sec:marl}
The problem of policy overfitting occurs as the estimated advantage is the same for all agents leading to the lack of credit assignment as described in Section \ref{sec:MAPPO}. Theoretically, this can be solved by ensuring the shared advantage value does not affect other agents by decomposing the centralized advantage value for each agent. However, given the nature of multi-agent systems, this is impractical. Therefore, Gaussian noise can be introduced as a regularization technique to reduce bias in the advantage values. This approach has been widely adopted in various machine learning applications for its simplicity and effectiveness in regularization \cite{goodfellow2016deep, igl2019generalization}. 

\begin{figure}
  \centering
  \begin{subfigure}[b]{\textwidth}
    \centering
    \includegraphics[width=0.7\textwidth]{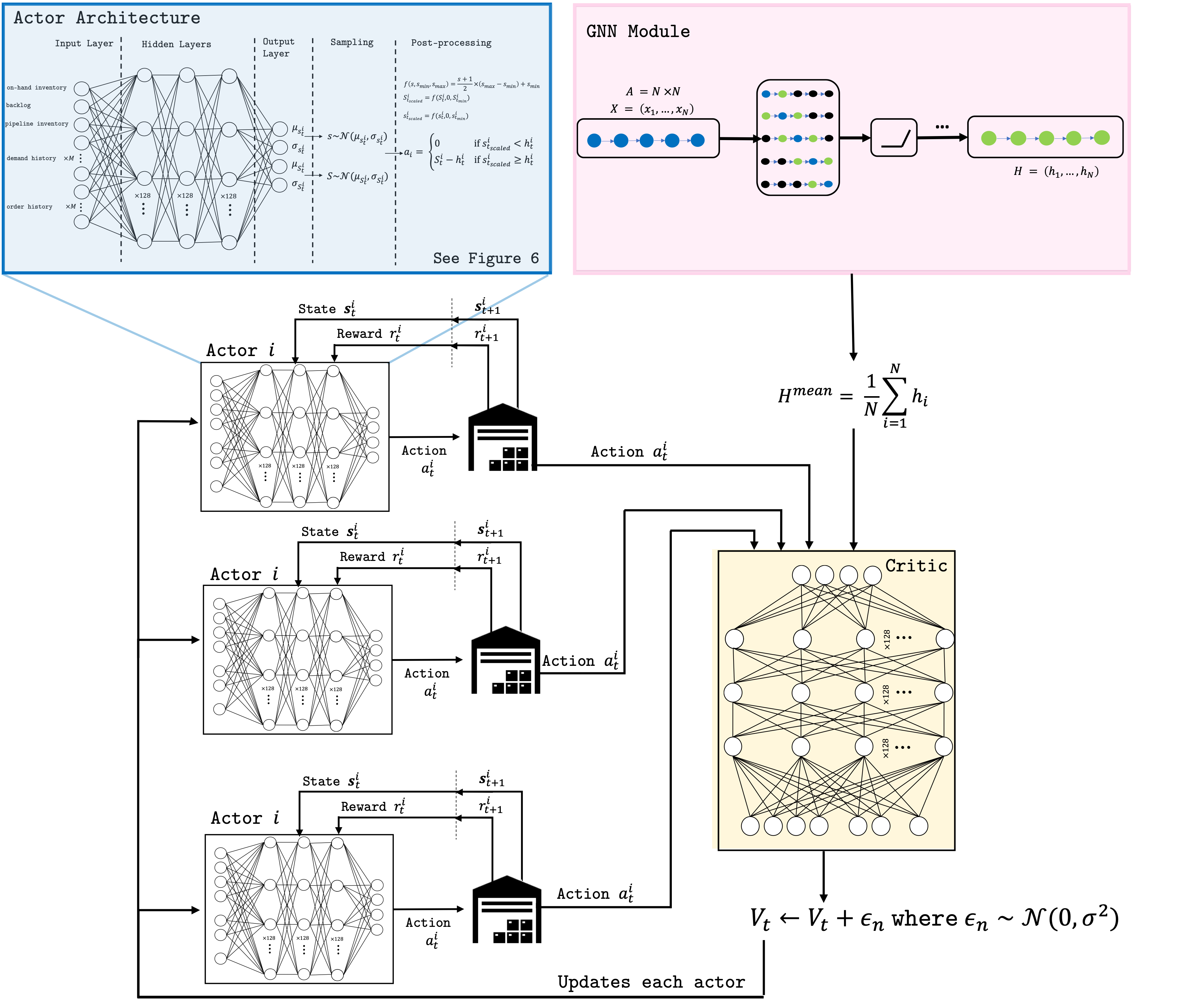}
    \caption{Overview of the training phase, illustrating the collaboration between the actor, critic, and graph convolutional neural networks for multiple actors.}
    \label{fig:final_train}
  \end{subfigure}
    \vskip\baselineskip 
  \begin{subfigure}[b]{0.7\textwidth}
    \centering
    \includegraphics[width=\textwidth]{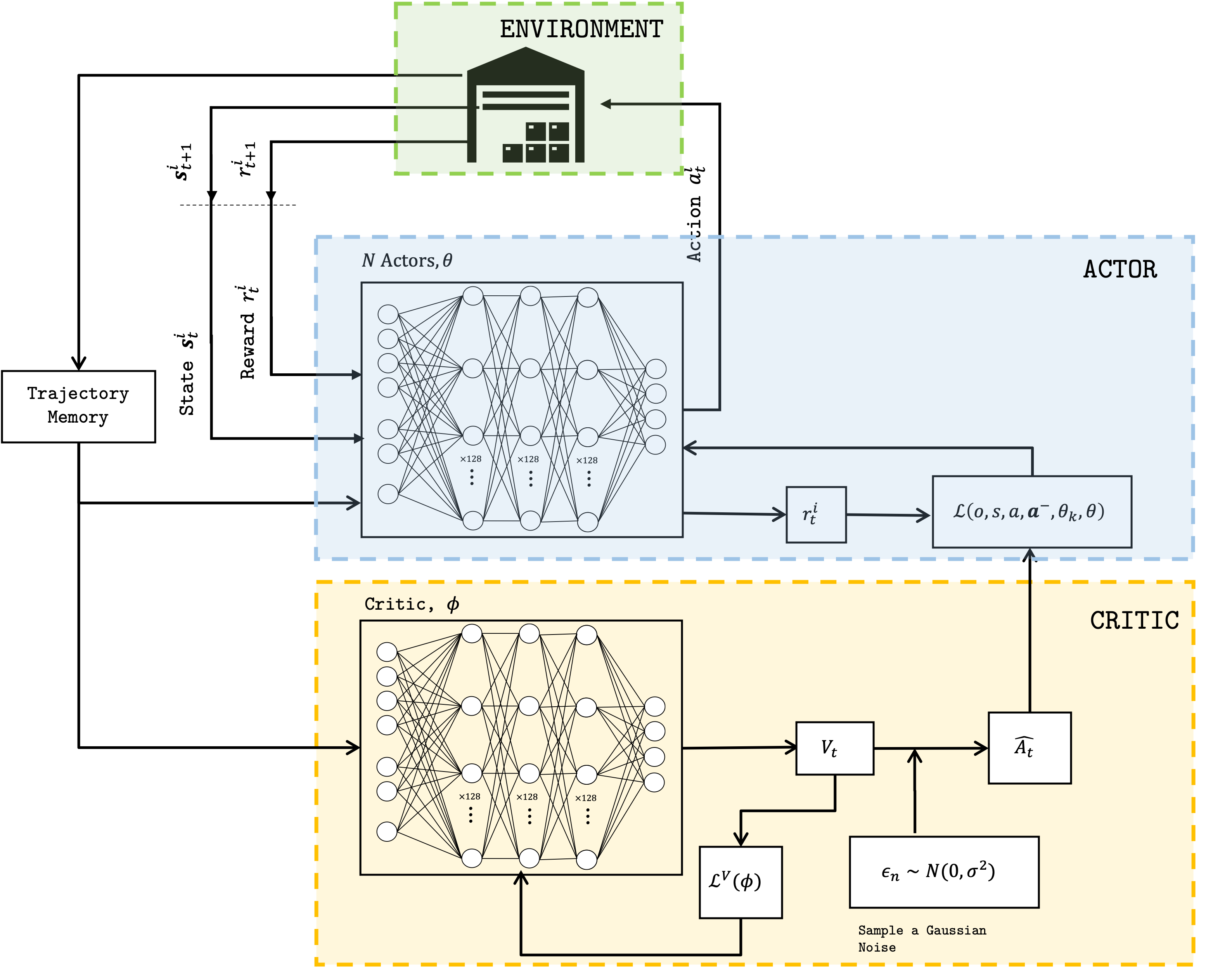}
    \caption{Illustration of value function noise injection within the actor-critic framework for a single actor.}
  \end{subfigure}
  \caption{Visual representations of the training processes: (a) the interaction among multiple actors during training, and (b) the role of noise injection in enhancing the learning dynamics shown with a single actor.} 
  \label{figure:noise-gp-mappo}

\end{figure}

Noise has been extensively studied and used in several forms across different components of reinforcement learning, including action space exploration and observation perturbation. In this work, we focus on introducing noise specifically in the value function which then propagates to the estimation of advantage values as shown in Figure \ref{figure:noise-gp-mappo}.  It is important to note that the framework shown in Figure \ref{figure:noise-gp-mappo} follows the same methodology described in Section \ref{sec:gcn_method}, with the key distinction being the incorporation of noise in the value function. This approach offers several potential benefits and implications: 
\begin{enumerate}
    \item \textbf{Reduces Overfitting and Biases.} Introduction of noise reduces over-fitting and biases in advantage value estimations. The introduction of randomness into the value function, means any patterns or biases that may arise from the agent's limited experience or observations are disrupted. 
    \item \textbf{Exploration Enhancement.} The introduction of noise in the value function promotes exploration by injecting randomness into the agent's value estimates.
\end{enumerate}
We sample a Gaussian noise, 
\begin{equation}
    \epsilon_n \sim \mathcal{N}(0, \sigma^2)
\end{equation}
where $\sigma^2$ is the variance (intensity) of the noise added to the samples. The value of $\sigma$ is a hyperparameter controlling the noise level, we analyse the effect of this hyperparameter in Section \ref{sec:sens_noise}. The global state, $s$, is then inputted into the centralized value network $V_{\phi}$ to which the sampled noise, $\epsilon_n$, is added. 
\begin{equation}
    V_{\phi}(s) \leftarrow V_{\phi}(s) + \epsilon_n
\end{equation}

The addition of the random noise disturbs the value function uniformly for all agents which is propagated to the advantage function. Uniform noise injection introduces variability into the advantage values, promoting exploration and adaptability across the entire agent population. This in turn provides benefits such as preventing overfitting caused by sampled advantage values with deviations and environmental non-stationarity. However, it's important to note that this approach may sacrifice the potential for agents to fully exploit their individual characteristics and preferences, which tailored noise could accommodate.

An additional advantage of Proximal Policy Optimization (PPO), is the inherent clipping mechanism of the algorithm which helps mitigate potential instability caused by the noise. The clipping ensures policy updates remain close to the previous policy allowing for exploration benefits introduced through value function noise injection without compromising training stability.

\section{Results and Discussion} \label{sec:r&d}
In this section four case studies of different inventory management configurations are used to illustrate the effectiveness of our proposed methodology as shown in Figure \ref{fig:supply_chain_networks}.  Detailed descriptions of these supply chains can be found in  \ref{appendix:sc}. We compare our proposed methodology (Reg-P-GCN-MAPPO and P-GCN-MAPPO) against other MARL methods including: IPPO, MAPPO, and Graph-based-MAPPO(G-MAPPO). This allows us to compare against state-of-the-art as well as analyze the contribution of the different components of our method. This is further benchmarked against single agent RL (PPO specifically) and a heuristic (s,S) policy. In the later, a static heuristic policy is found for each node in the network, where optimal parameters are found using a derivative-free method, implemented using SciPy \citep{scipy} with a multi-start approach combined with local search. 

All variant MARL algorithms are implemented in the Ray RLLib framework \citep{rllib}, and all the implementation details including neural network architectures and hyperparameters can be found in  \ref{appendix:NN} and \ref{appendix:hyp}.

\begin{figure}
    \centering
    \includegraphics[width=0.8\textwidth]{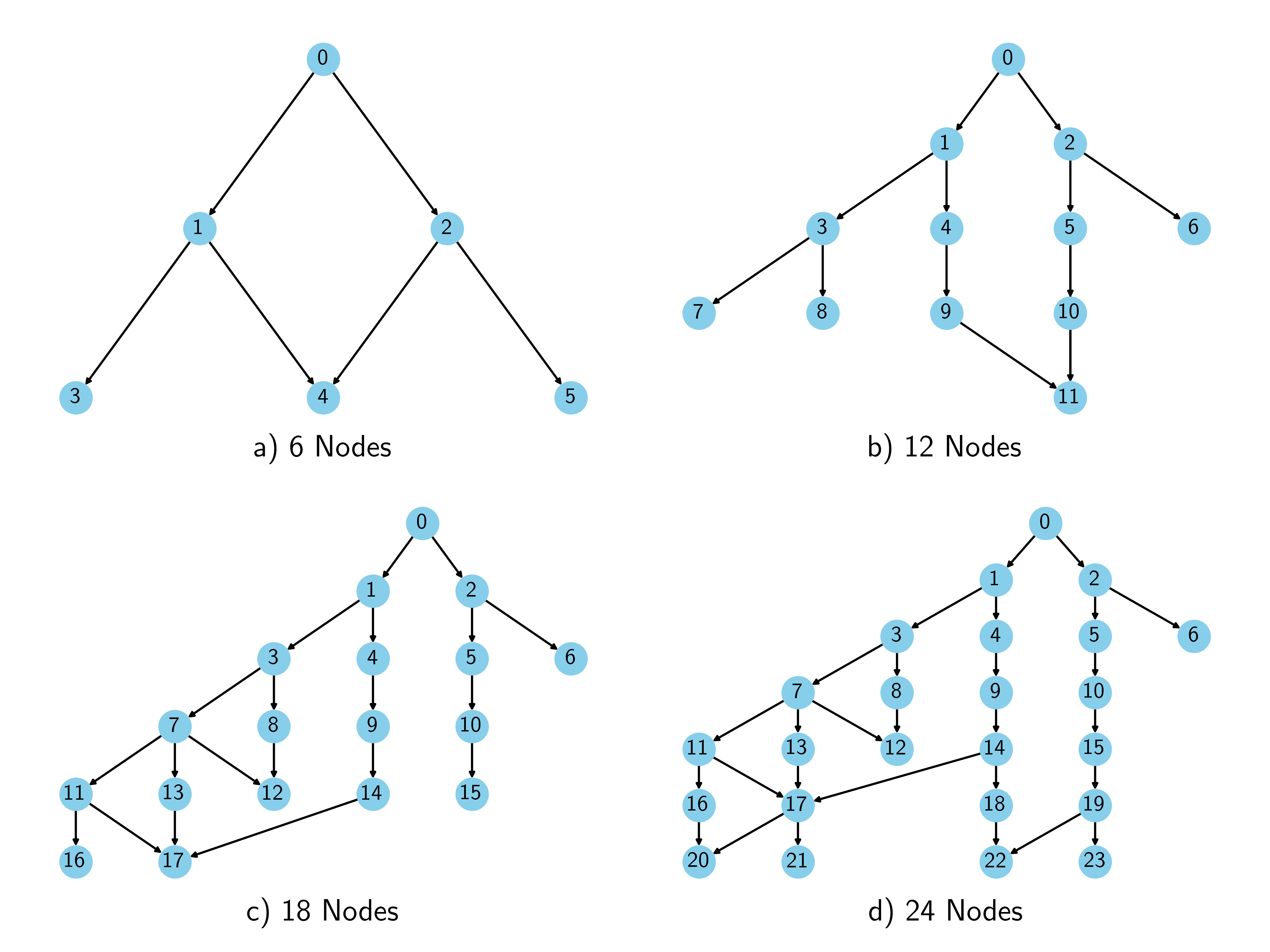}
    \caption{Different supply chain configuration case studies with varying number of nodes}
    \label{fig:supply_chain_networks}
\end{figure}

\subsection{Execution}
To evaluate the performance and robustness of our proposed methodology, we considered different scenarios involving a varying number of agents: 6, 12, 18, and 24 agents. Each scenario represents a different level of complexity and coordination required among the agents. In order to assess our methodology, we observe the performance of the trained policies on 20 simulated test episodes with 50 time-steps each. Both the lead time and demand uncertainty remained constant throughout the test episodes and each methodology. All methodologies follow the CTDE framework where agents only require local information at execution. The execution of the MARL algorithms in these scenarios was assessed through performance metrics including cumulative reward, backlog and inventory levels.

\begin{figure}
    \centering
    \includegraphics[width = 0.7\textwidth]{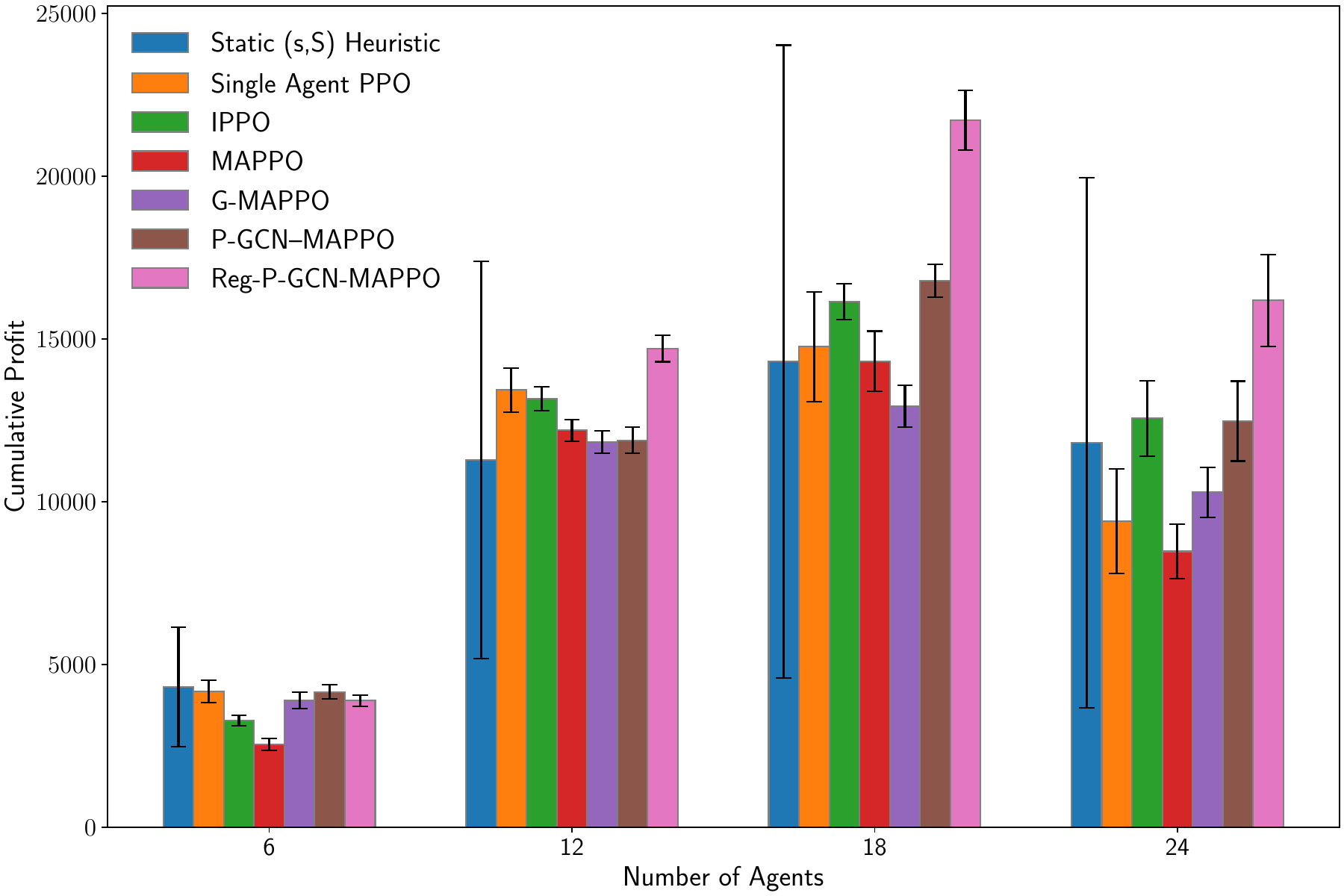}
    \caption{Cumulative Profit (with Standard Deviation) for the different methodologies across the four different supply chain configurations}
    \label{fig:profit}
\end{figure}

Firstly, the bar graph shown in Figure \ref{fig:profit} compares the cumulative profit for the different methodologies in the different supply chain network configurations. Notably, the proposed methodology, Reg-P-GCN-MAPPO, has on par performance or outperforms the other methodologies. This advantage becomes increasingly pronounced as the number of agents rises. This trend suggests that Reg-P-GCN-MAPPO may be particularly effective in large-scale, multi-agent inventory control scenarios. The decrease in cumulative profit from the 18 agent to 24 agent configuration can be attributed to differences in number of retailers present and the specific environment configuration chosen. Figure \ref{fig:profit} also shows that while the static $(s_{t_\text{inv,s}}^{i}, S_{t_\text{ord,s}}^{i})$ policy achieves similar cumulative profit performance to Reg-P-GCN-MAPPO, it exhibits a significantly higher standard deviation. This variability suggests that the static heuristic is less stable, potentially hindering its robustness in situations with unexpected changes to the environment. 

This distinction is crucial in real-world applications, where supply chain networks and other multi-agent systems inevitably encounter unexpected events. RL's ability to maintain good performance in uncertain environments is one of the key advantages compared to traditional methods.  Therefore, the results highlight the strength of RL methods, particularly Reg-P-GCN-MAPPO, as its lower standard deviation suggests greater potential for robustness and adaptability in uncertain environments.

\begin{figure}
  \begin{subfigure}[b]{0.5\textwidth}
    \centering
    \includegraphics[width=\textwidth]{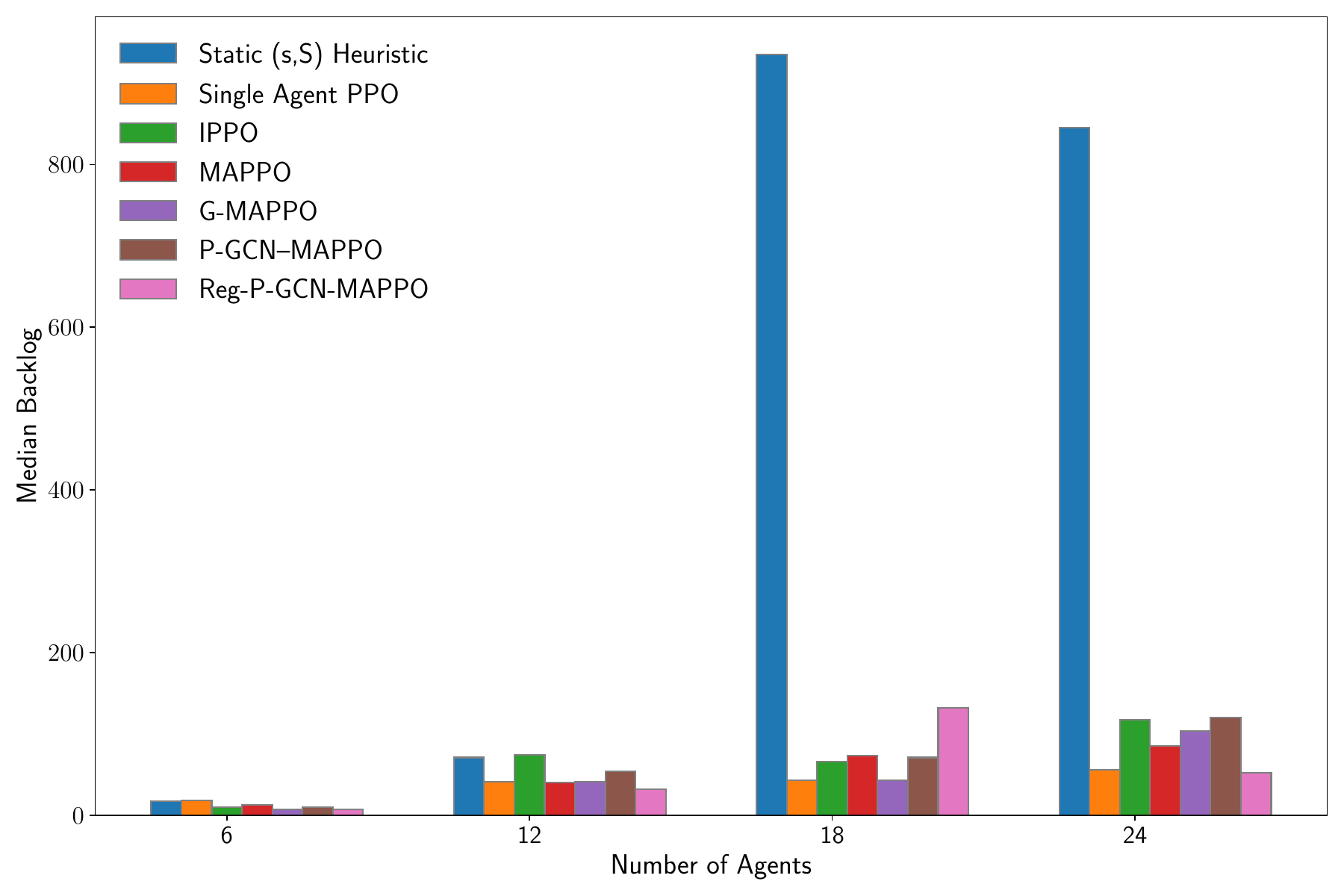}
    \caption{Backlog}
    \label{fig:backlog}
  \end{subfigure}
  \hfill
  \begin{subfigure}[b]{0.5\textwidth}
    \centering
    \includegraphics[width=\textwidth]{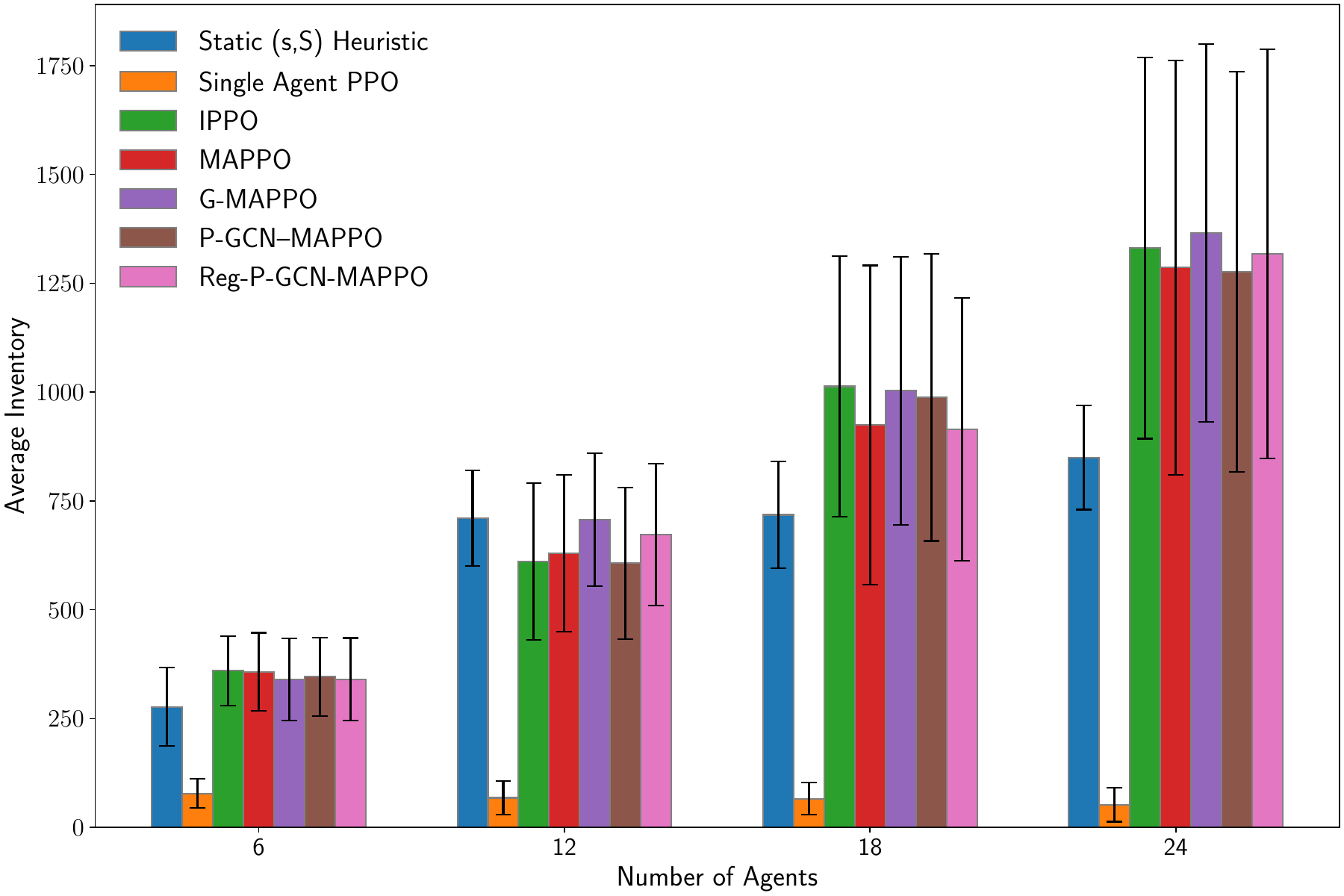}
    \caption{Inventory}
    \label{fig:inventory}
  \end{subfigure}
  \caption{Median metric along with standard deviations for the different methodologies and configurations}  
\end{figure}

Figure \ref{fig:backlog}, illustrates the median backlog for different algorithms. The static $(s_{t_\text{inv,s}}^{i}, S_{t_\text{ord,s}}^{i})$ heuristic exhibits a significant increase in backlog as the number of agents rises. This trend suggests potential difficulties in handling the growing complexity of the environment with more agents (e.g., increased number of interactions, coordination challenges). The static nature of the inventory policy in the static $(s_{t_\text{inv,s}}^{i}, S_{t_\text{ord,s}}^{i})$ heuristic might also contribute, as it cannot adapt to changing environmental conditions. In contrast, our proposed methodology addresses this limitation by redefining the action space to parametrize a heuristic policy. This allows our policy to be more adaptive and adjust its behavior based on the current state of the environment, potentially leading to better handling of increased complexity and reducing the backlog. However, from Figure \ref{fig:inventory}, only minimal differences were present in the final on-hand inventory across methodologies.

Our methodology is also compared with other MARL methods. The notable difference between them is the information available to the critic. One key challenge in MARL is balancing access to global information (mitigating non-stationarity) with scalability limitations of naively combining all agent states. Our proposed methodologies, P-G-MAPPO and Reg-P-GCN-MAPPO, address this by leveraging graph-based approaches and the inherent structural properties of the problem. This allows agents to exploit both local and global information effectively.

\begin{figure} 
  \centering
  \begin{subfigure}[b]{0.475\textwidth}
    \centering
    \includegraphics[width=\textwidth]{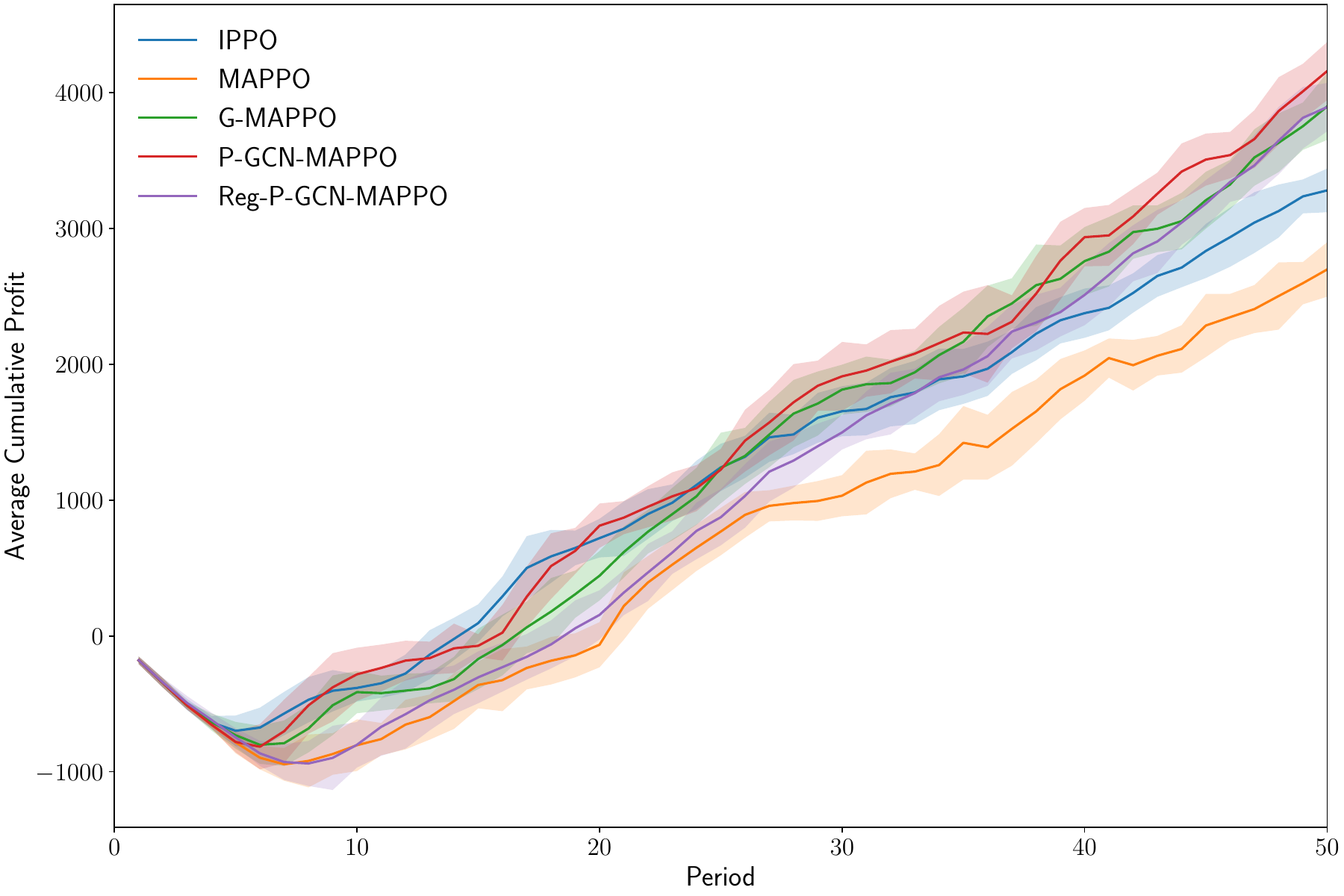}
    \caption{6 nodes}
  \end{subfigure}
  \hfill
  \begin{subfigure}[b]{0.475\textwidth}
    \centering
    \includegraphics[width=\textwidth]{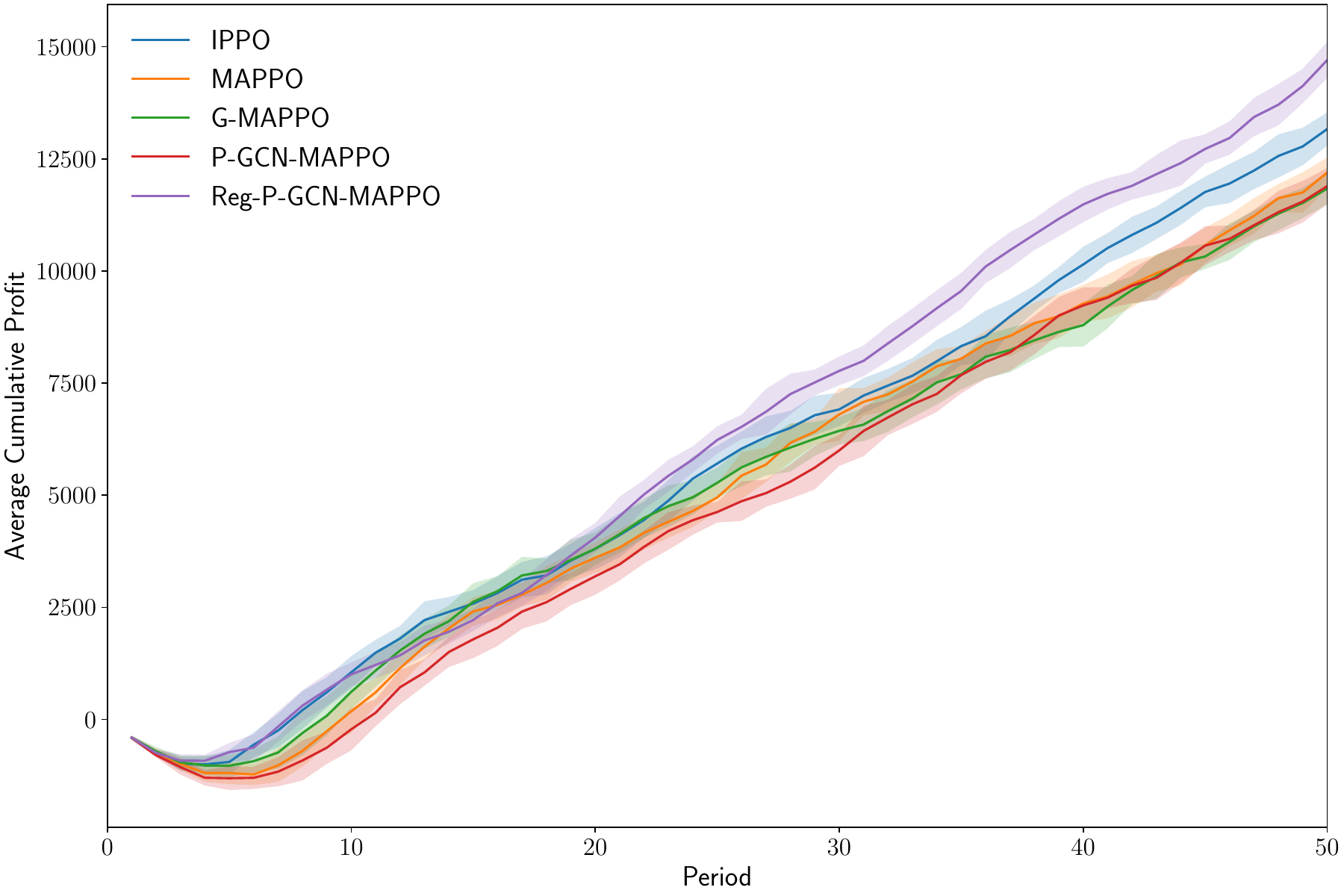}
    \caption{12 nodes}
  \end{subfigure}
  \vskip\baselineskip
  \begin{subfigure}[b]{0.475\textwidth}
    \centering
    \includegraphics[width=\textwidth]{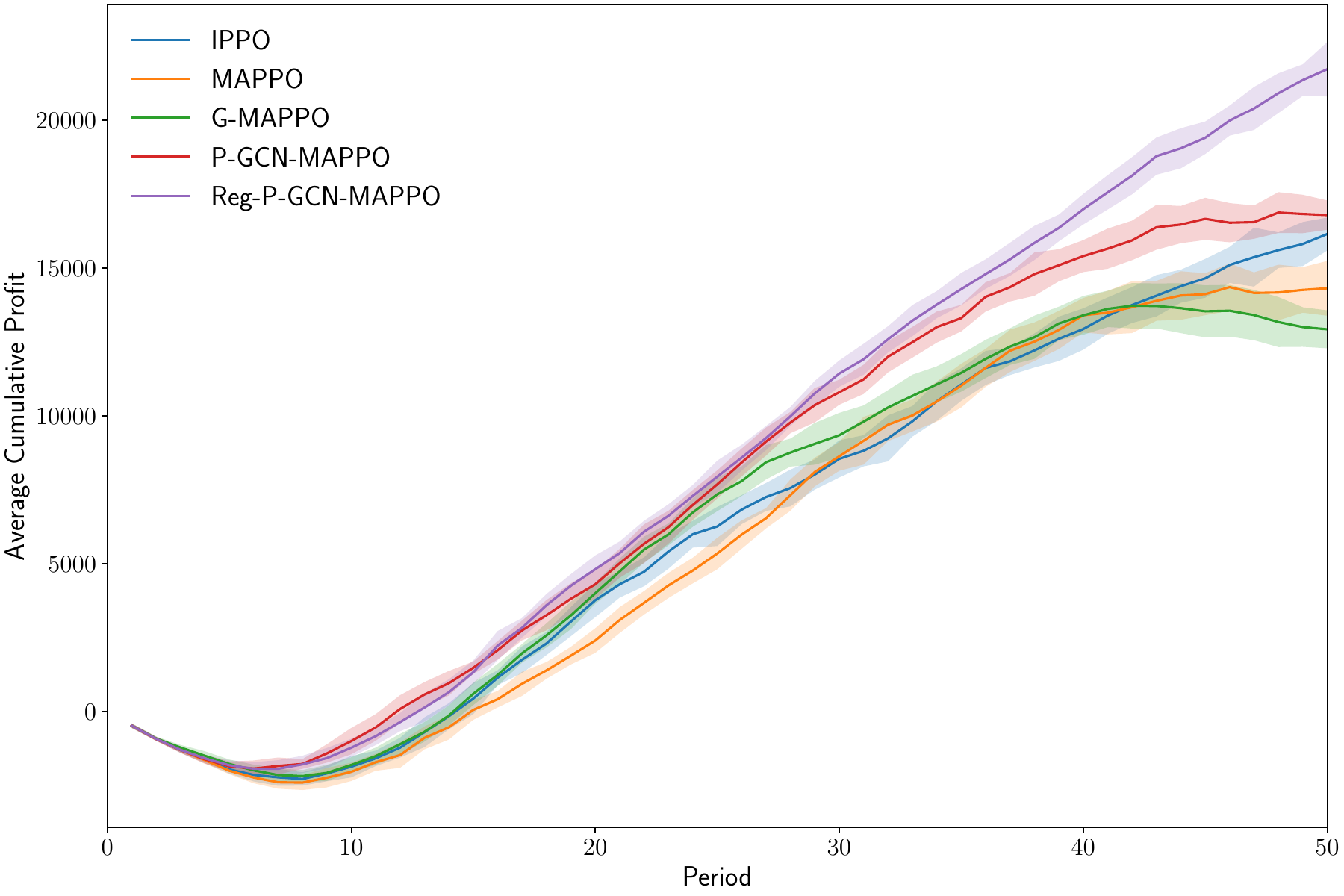}
    \caption{18 nodes}
  \end{subfigure}
  \hfill
  \begin{subfigure}[b]{0.475\textwidth}
    \centering
    \includegraphics[width=\textwidth]{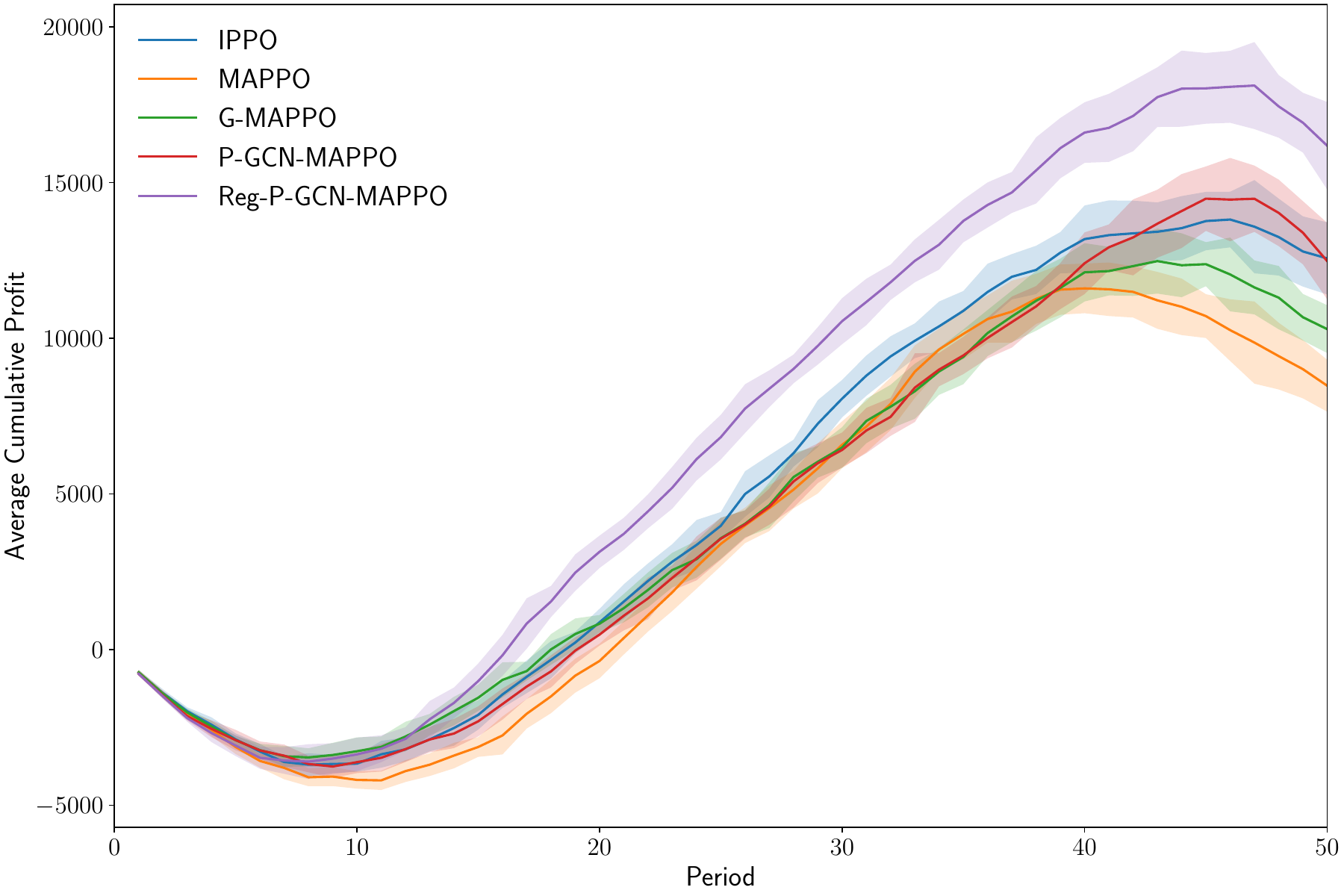}
    \caption{24 nodes}
  \end{subfigure}
  \caption{Execution curves for the 4 different supply chain configurations.} 
  \label{fig:execute}
\end{figure}

Figure \ref{fig:execute} shows how IPPO consistently outperforms MAPPO, and this performance gap widens with a greater number of agents. This motivated the development of smarter methods of information aggregation, by leveraging graph-based approaches, where information can be harnessed whilst not suffering from performance shortcomings. This highlights the need for smarter information aggregation. G-MAPPO, P-GCN-MAPPO and Reg-P-GCN-MAPPO incorporate Graph Neural Networks (GNNs) to enable communication and information sharing while maintaining focus on local details. 

However, from Figure \ref{fig:execute}, the G-MAPPO method does not always outperform MAPPO and IPPO methods. This can be attributed to the excessive capture of information, leading to overfitting of the policies. Specifically, when too much information is incorporated, the policies tend to overfit the training data. This is consistent with prior research, where \citet{nayak2023scalable} highlighted how excessive information aggregation in multi-agent systems without proper regularization can lead to overfitting, stressing the importance of intelligent information aggregation strategies to scale reinforcement learning methods effectively. \citet{wang2022noise} also showed MAPPO may suffer from insufficient exploration due to its reliance on batch-sampled experiences from a replay buffer, which can lead to policy overfitting   In contrast, in environments where exploration is critical, IPPO's independent nature may facilitate more effective exploration strategies, allowing agents to discover diverse policies that are less prone to overfitting \citep{wang2021multi}.

In  light of this, P-GCN-MAPPO outperforms G-MAPPO, MAPPO and IPPO in all 4 case studies. This occurs as the integration of a pooling mechanism, can help aggregate information from different parts of the graph, reducing the dimensionality of the data acting as a regularizer and potentially reducing overfitting. The message passing inherent in graph neural networks allows for agents to inherently communicate with each other whilst the pooling mechanism prevents overfitting of the policies. Despite losing local information in the critic, the global mean pooling mechanism allows for global information to be captured in the central critic that helps reduce the variance in the individual actors that are trained with local information.

Our final proposed methodology hypothesizes that the addition of noise into the value function, propagates variability into the advantage function potentially reducing overfitting. The execution curves shown in Figure \ref{fig:execute} show the impact of adding noise to the value function on our methodology’s performance. The results show that with a small number of agents (6 agents), adding noise into the value function does not outperform P-GCN-MAPPO compared to when the number of agents is larger (24 agents). When the number of agents is 6, the overall state space is smaller compared to a large agent system. In larger agent systems, the addition of noise can also be seen as regularization, where the policies become more robust to the complex environment. 

\begin{figure}
    \centering
    \includegraphics[width=0.5\textwidth]{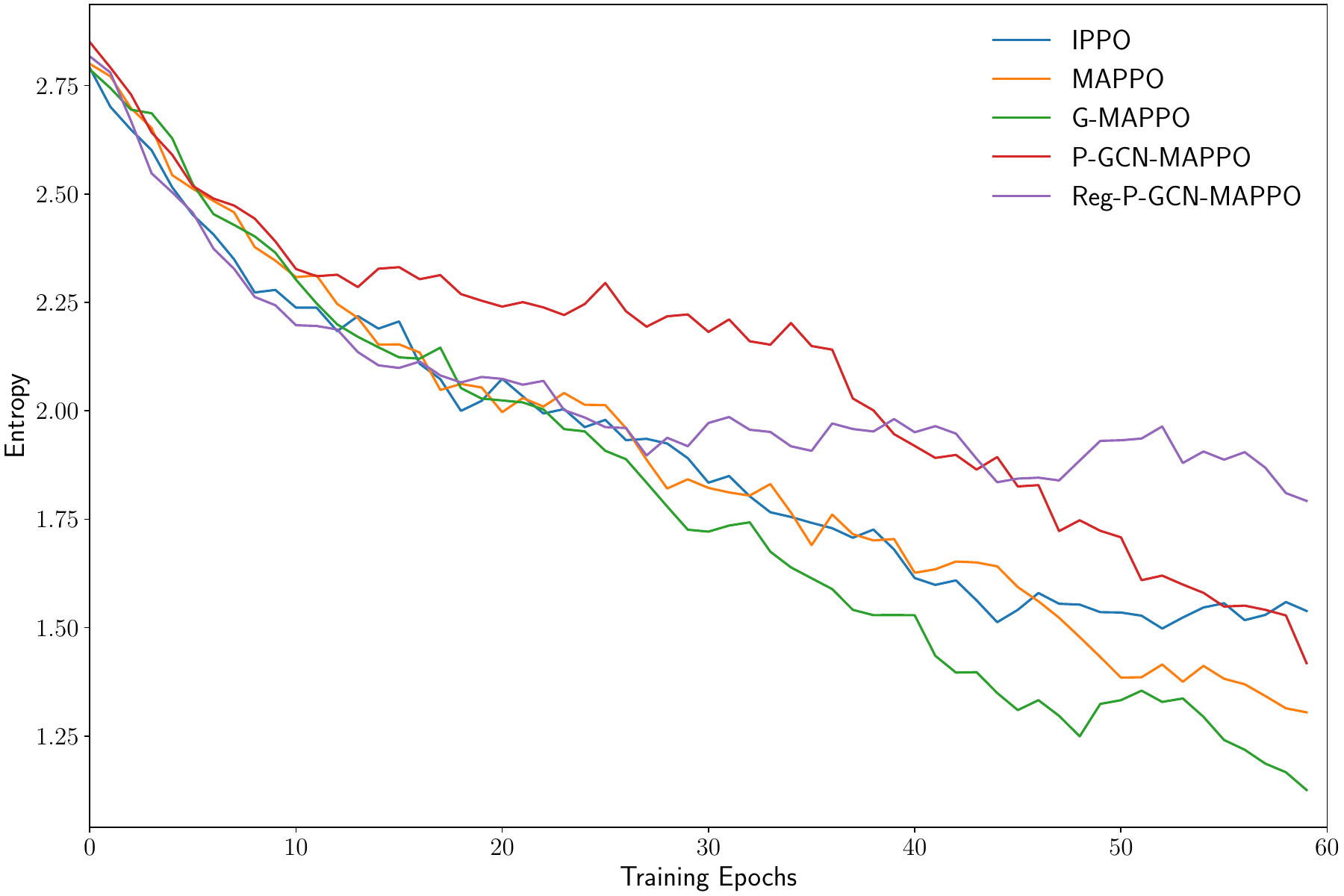}
    \caption{Entropy, as defined in reinforcement learning, represents the uncertainty or randomness in the policy's action distribution \citep{sutton2018reinforcement}. The figure shows the change in entropy during the training process for the different methods in the 18 agent configuration}
    \label{fig:entropy18agents}
\end{figure}

Moreover, it is useful to compare the change in entropy during the training phase for each of the different methods. In the context of MARL, entropy quantifies the uncertainty or randomness in the policy's action distribution and is calculated based on the RL definition as the negative expectation of the log probabilities of the policy's action distribution \citep{sutton2018reinforcement}. During the training process, as the agents learn, the entropy in the policy distribution decreases as the agents learn better policies. Figure \ref{fig:entropy18agents} shows that the rate of entropy decrease differs between the different methods. The addition of noise to the value function introduces controlled variability which propagates into the advantage estimates. This helps maintain a higher entropy for a longer period during training which is desirable as high entropy promotes exploration. During MARL training, agents need to balance exploring new actions and exploiting known good actions. By exploring a wider range of actions (due to high entropy), agents are more likely to discover better strategies and avoid getting stuck in sub-optimal solutions. This can be especially important in complex environments where the best course of action depends on what other agents are doing. Figure \ref{fig:entropy18agents} shows that Reg-P-GCN-MAPPO, maintains a higher entropy throughout training, promoting exploration and reducing policy overfitting leading to superior performance. Such behavior is particularly advantageous in environments characterized by stochasticity, where partial observability, randomness in rewards or transitions, and non-stationary dynamics are present \citep{su2022divergence}. In these settings, a policy with higher entropy is more likely to remain flexible, allowing agents to adapt to the evolving state of the environment. We hypothesize the persistence of higher entropy even after convergence is from the stochasticity introduced by entropy regularization, which leads the agents' policies to stabilize around distributions that retain a degree of randomness. This is particularly crucial in cooperative scenarios, where agents must balance individual and collective goals. The added complexity of multiple interacting agents necessitates that policies remain responsive to the actions of others, which can manifest as sustained entropy within the policy distributions \citep{su2022divergence}.

In MARL environments, the non-stationary nature of the system, driven by the simultaneous learning and adaptation of multiple agents, further highlights the need for flexible policies. Each agent's policy can influence the others, leading to a continuously changing environment. Policies that maintain higher entropy enable agents to remain adaptable, preventing them from becoming overly specialized or rigid in their strategies. This flexibility is crucial for sustaining performance in environments where the actions of other agents significantly impact the state dynamics \citep{nguyen2020deep, li2020parallel}. Thus, while traditional methods may favor policies with lower entropy—indicative of more deterministic, greedy behaviors—the benefits of maintaining higher entropy are especially apparent in dynamic, multi-agent settings, where adaptability and flexibility are key to successful coordination and performance.

\subsection{Sensitivity to Gaussian Noise} \label{sec:sens_noise}
In addition to acting as a regularizer, the incorporation of Gaussian Noise into the value function of MARL algorithms facilitates a balance between effective exploration and exploitation, a well known phenomena in the field of reinforcement learning. The following section will explore this sensitivity to noise intensity, examining how both insufficient and excessive noise can negatively impact the performance of MARL agents. 

\begin{figure}
  \begin{subfigure}[b]{0.5\textwidth}
    \centering
    \includegraphics[width=\textwidth]{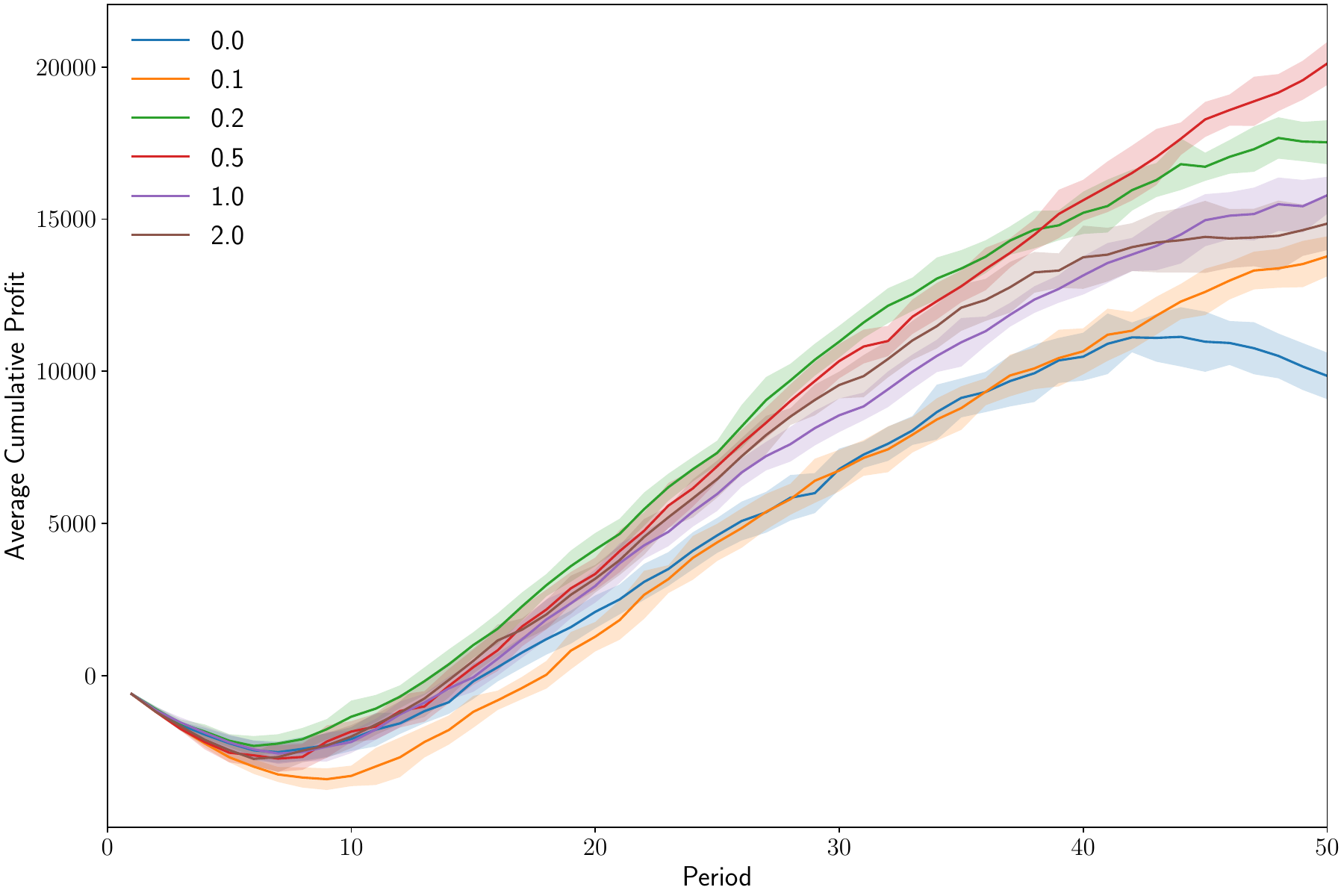}
    \caption{Execution curves after training. The mean episode reward is shown across 20 evaluation episodes, with shaded standard deviation.}
    \label{fig:execute_noise}
  \end{subfigure}
  \hfill
  \begin{subfigure}[b]{0.5\textwidth}
    \centering
    \includegraphics[width=\textwidth]{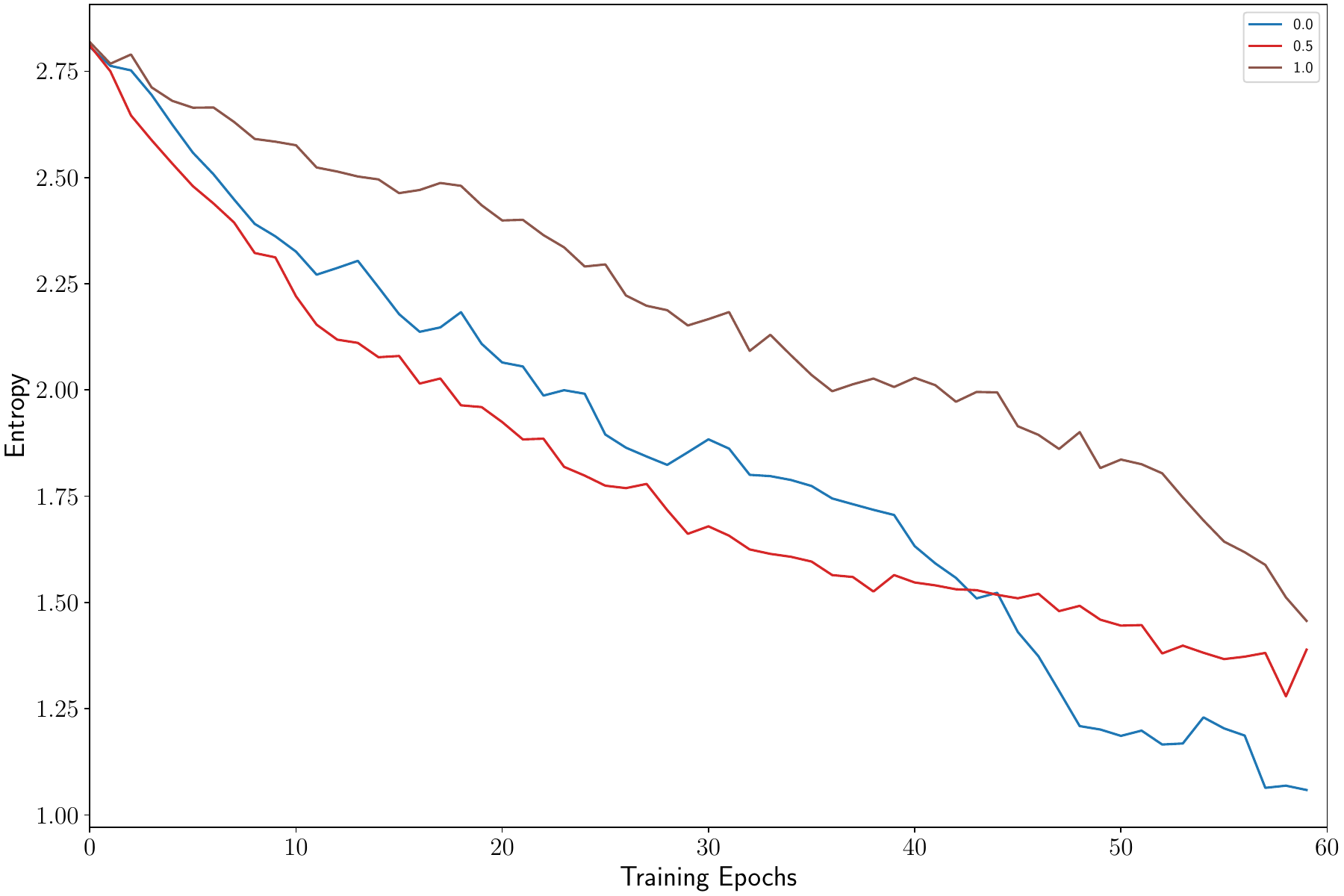}
    \caption{Entropy change during training for three different noise levels, illustrating the impact of noise intensity on policy exploration}
    \label{fig:entropy_noise}
  \end{subfigure}
  \caption{Sensitivity of noise intensity was performed for the 18 agent configuration}  
\end{figure}

The standard deviation values were changed from 0.0, 0.1, 0.2, 0.5, 1.0 and 2.0. To evaluate the impact of noise intensity, we compare the mean reward across 20 evaluation episodes, with the standard deviation shown as a shaded region in Figure 12a. This helps assess both policy performance and stability across different noise levels. The performance increases up to a standard deviation of 0.5. After which, when the noise intensity increases, the overall performance decreases. This occurs as when the standard deviation of the Gaussian distribution is small, there's insufficient noise, leading to a lack of regularization. This can result in overfitting, limited exploration, and a tendency to settle into local optima. This is particularly detrimental in scenarios with large state spaces, where effective exploration becomes crucial for agents to learn optimal policies. Conversely, when the standard deviation is 1.0 and 2.0 the overall performance becomes worse as seen in Figure \ref{fig:execute_noise}. This occurs when the noise intensity is too high, excessive noise can introduce significant randomness, increasing the variance in the value function which disrupts the learning process which may hinder convergence. 

Moreover, the entropy during training is compared in Figure \ref{fig:entropy_noise} for the noise value that performed best [0.5], the noise value that performed worst [0.1] and the base case scenario [0.0]. The results show that using a noise level with standard deviation = 0.5 leads to a higher final entropy compared to the scenario with no noise (standard deviation = 0.0). However, throughout the training process, when the noise intensity is high (standard deviation = 2.0), the entropy remains higher, emphasizing that this noise has introduced significant randomness leading to suboptimal policies. 

Therefore, finding a  level of noise becomes important, as it can act as a regularizer, promoting exploration in complex environments while maintaining a level of stability that allows for effective learning. This delicate balance is particularly important in settings with a large number of agents, where the vast state space and intricate interactions require a measured approach to exploration via noise injection. 

\subsection{Uncertainty}
In this section, we investigate how uncertainty affects the robustness of the proposed framework and its ability to adapt to the complexities of inventory management. Uncertainty, as defined by the International Bureau of Weights and Measures, represents the inherent doubt in measurement results, which has critical implications for decision-making and operational efficiency. In inventory management, uncertainty arises from sources such as demand variability, supply chain disruptions, and lead time fluctuations. These uncertainties can result in overstocking or stockouts, affecting profitability and operational reliability. 

\begin{figure}
    \centering
    \includegraphics[width=0.5\linewidth]{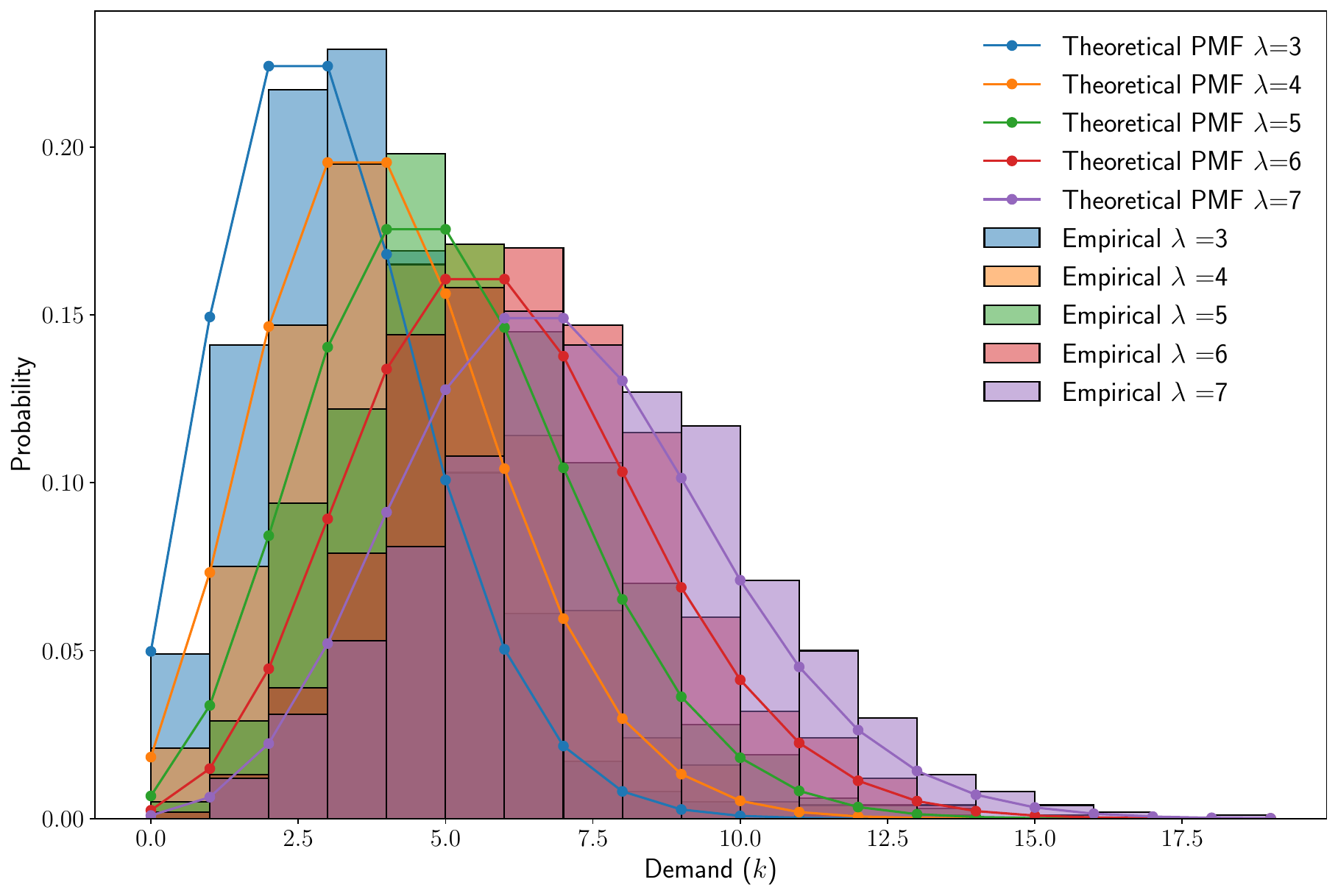}
    \caption{Comparison of empirical demand distributions with the theoretical Poisson PMF for different values of $\lambda_\text{d}$.}
    \label{fig:unceertain_demand}
\end{figure}

In our inventory management model, demand follows a Poisson distribution with a mean $\lambda_\text{d}$. The empirical demand distributions were compared against the theoretical Poisson Probability Mass Function (PMF) as shown in Figure \ref{fig:unceertain_demand}. The results show that demand uncertainty grows with $\lambda_\text{d}$, with higher values leading to an increased probability of extreme demand spikes. 

To assess the robustness of our learned RL policies, we trained the agents with our developed methodology (Section \ref{sec:marl}) under $\lambda_\text{d} = 5$ and tested it under varying demand conditions $\lambda_\text{d} = 3,4,5,6,7$. 

\begin{figure}
  \begin{subfigure}[b]{0.5\textwidth}
    \centering
    \includegraphics[width=\textwidth]{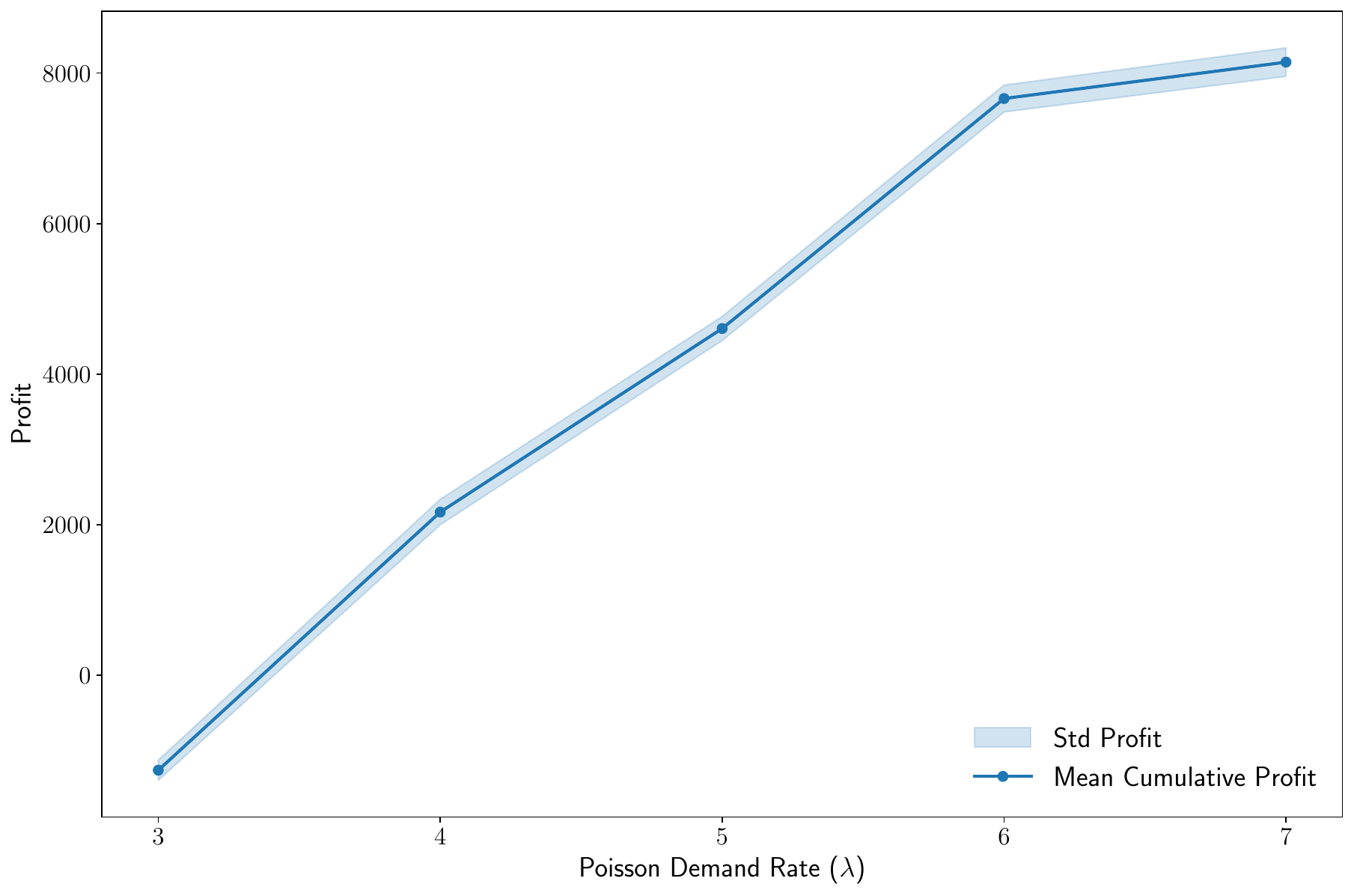}
    \caption{Profitability under varying demand levels}
    \label{fig:uncertain_profit}
  \end{subfigure}
  \hfill
  \begin{subfigure}[b]{0.5\textwidth}
    \centering
    \includegraphics[width=\textwidth]{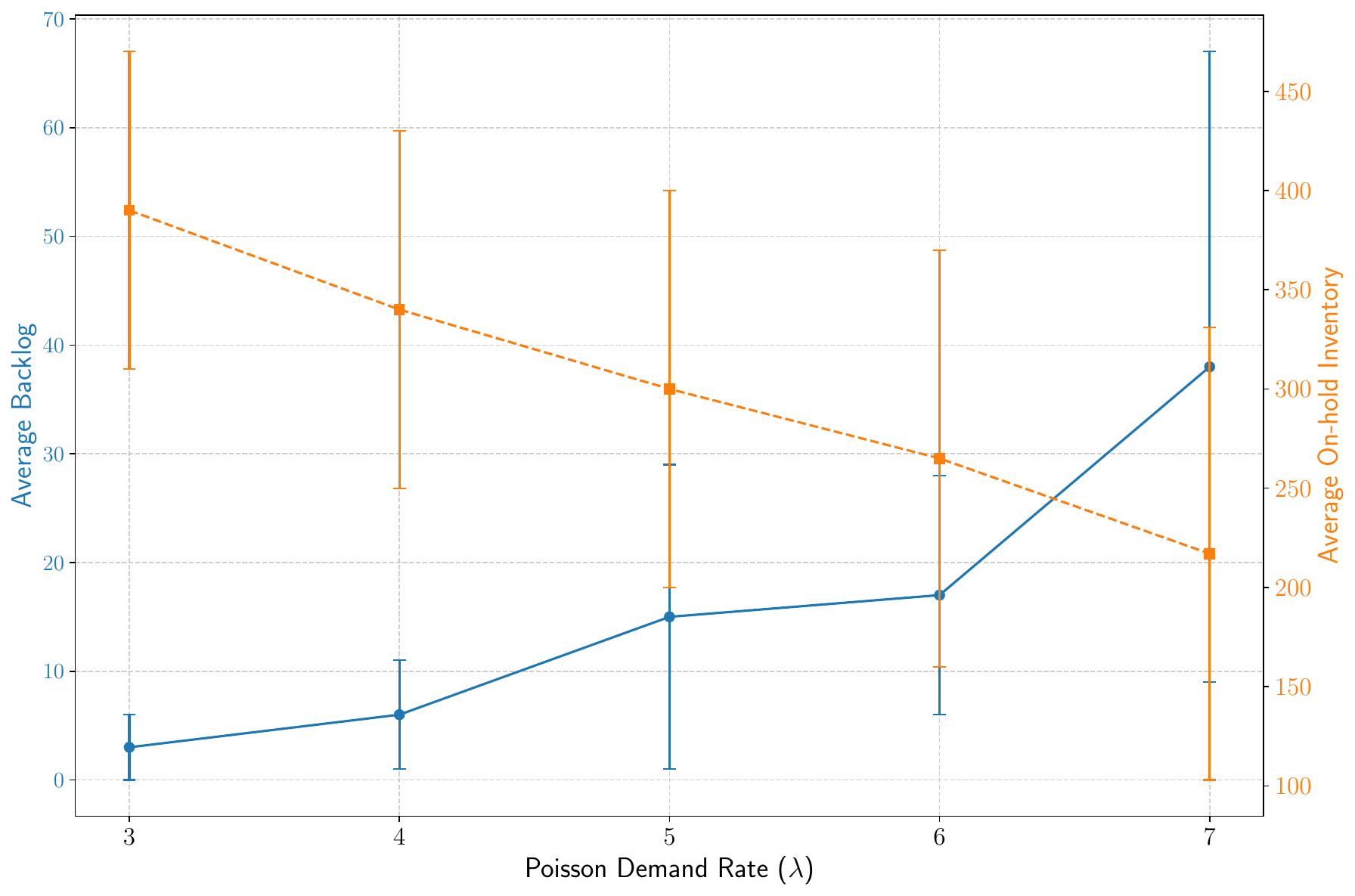}
    \caption{Inventory levels and backlog across demand levels}
    \label{fig:unc_inventory}
  \end{subfigure}
  \caption{Mean values with standard deviations for profitability, inventory levels, and backlog under different demand configurations.}
\end{figure}

Profitability is sensitive to demand fluctuations, as shown in Figure \ref{fig:uncertain_profit}. At low demand levels, the RL agent tends to overstock, resulting in suboptimal profits. However, as demand increases, the variance in profit remains relatively stable, suggesting the policy maintains robustness despite fluctuations.

As expected, backlog increases with higher $\lambda_\text{d}$ values, reflecting the agent's limited capacity to fulfill all demand within each time step. Conversely, on-hold inventory decreases with increasing demand, showing an inverse relationship with backlog, as shown in Figure \ref{fig:unc_inventory}.

As demand uncertainty grows, the standard deviation of inventory and backlog levels increases, indicating greater volatility in stock availability. Despite this, profit variance remains stable, suggesting that the RL policies are robust and maintain predictable financial performance. This suggests that the agent is effectively adapting its inventory and backlog management to maximize profits, despite increased demand variability.

\subsection{Scalability}
This section explores the scalability of the different methodologies examined in this paper. 

\begin{figure}
  \centering
  \begin{subfigure}[b]{0.475\textwidth}
    \centering
    \includegraphics[width=\textwidth]{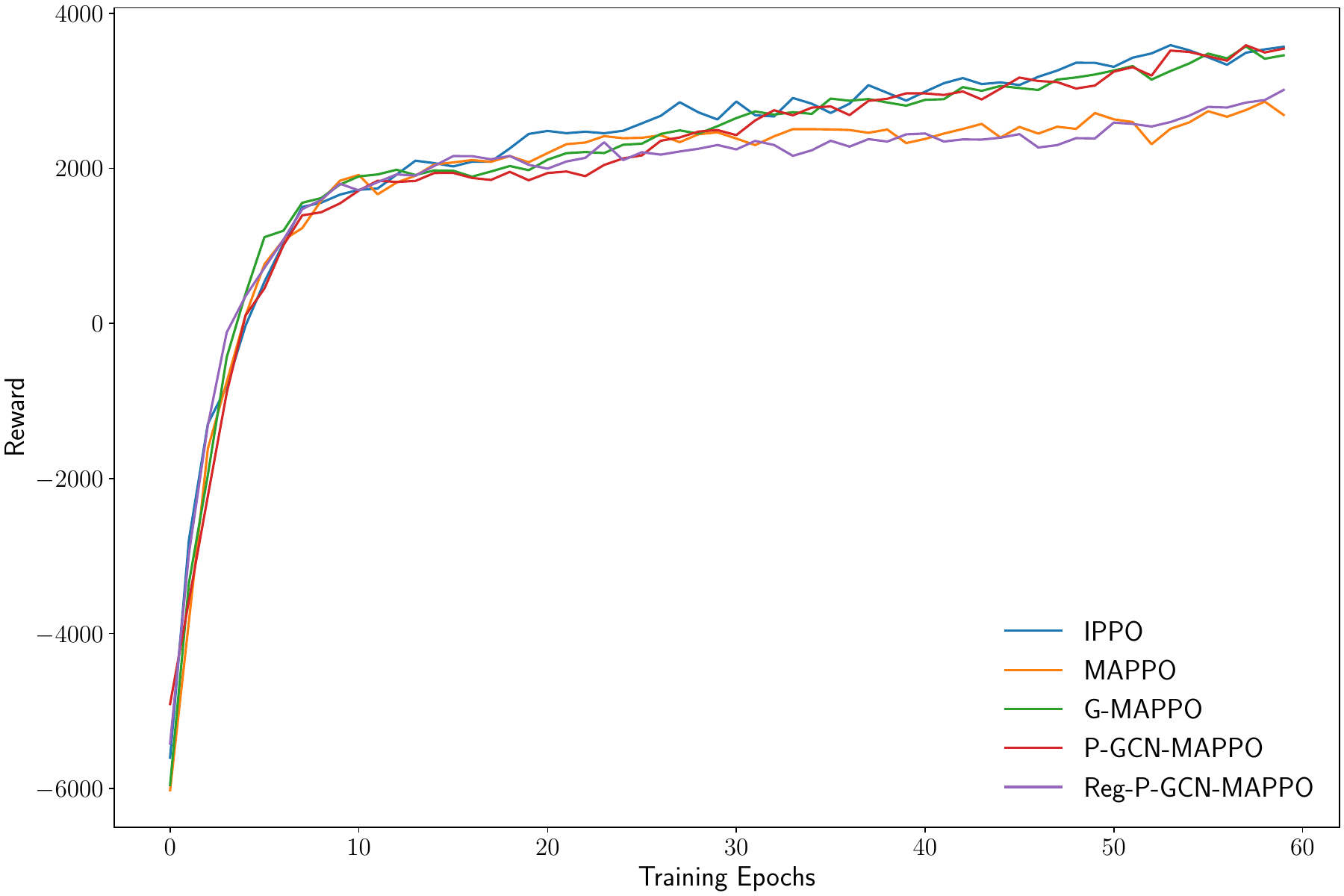}
    \caption{6 agents}
  \end{subfigure}
  \hfill
  \begin{subfigure}[b]{0.475\textwidth}
    \centering
    \includegraphics[width=\textwidth]{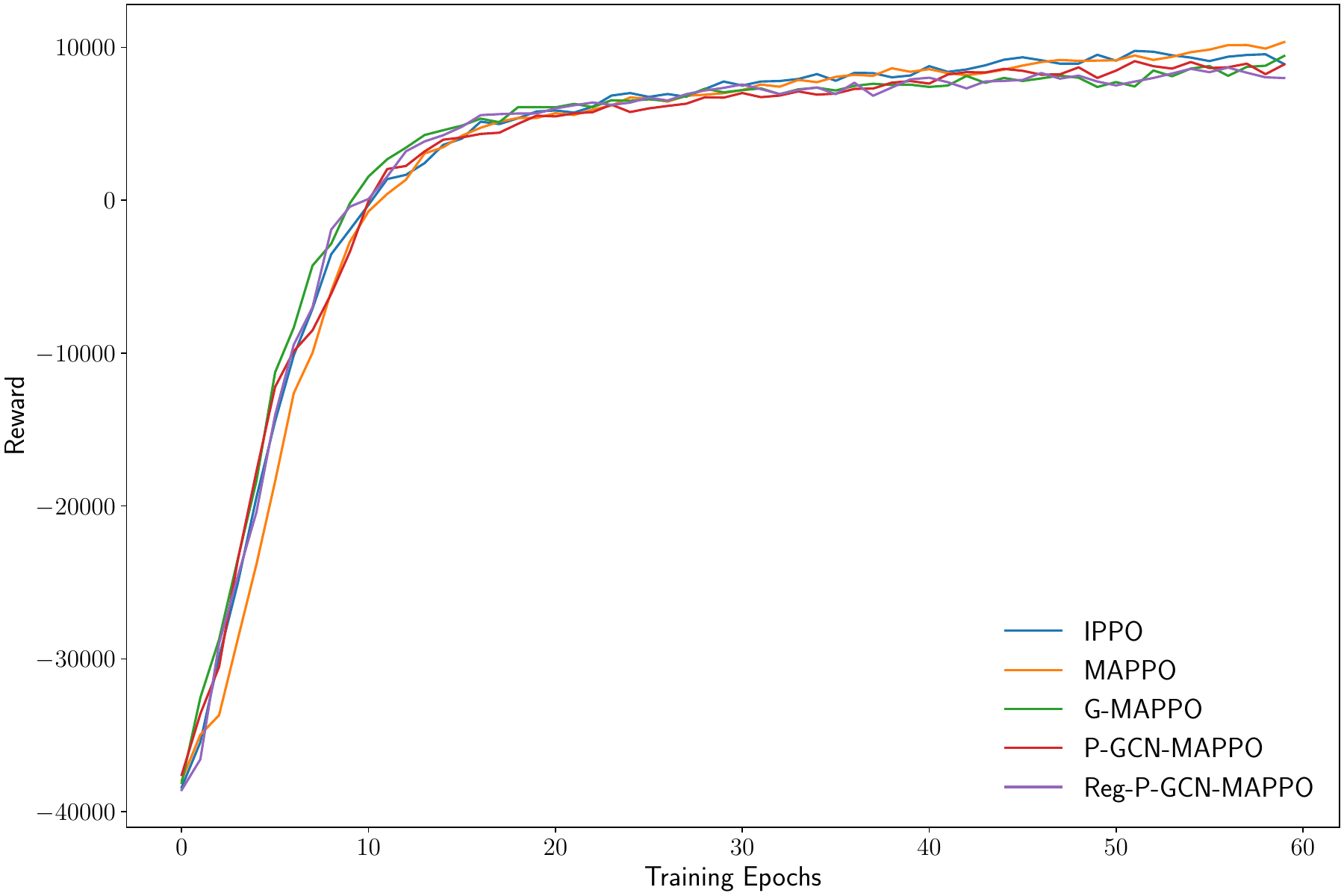}
    \caption{12 agents}
  \end{subfigure}
  \vskip\baselineskip
  \begin{subfigure}[b]{0.475\textwidth}
    \centering
    \includegraphics[width=\textwidth]{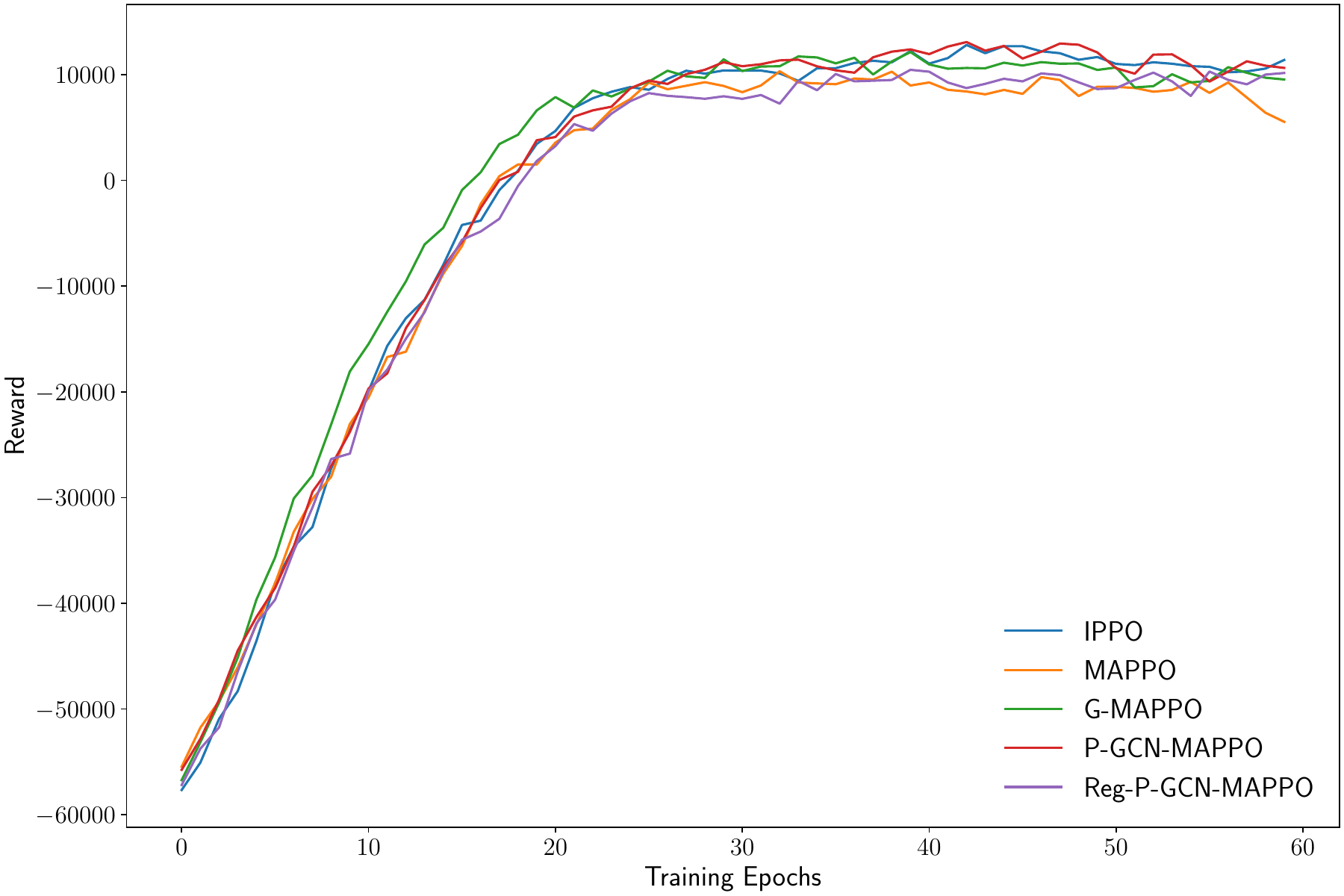}
    \caption{18 agents}
  \end{subfigure}
  \hfill
  \begin{subfigure}[b]{0.475\textwidth}
    \centering
    \includegraphics[width=\textwidth]{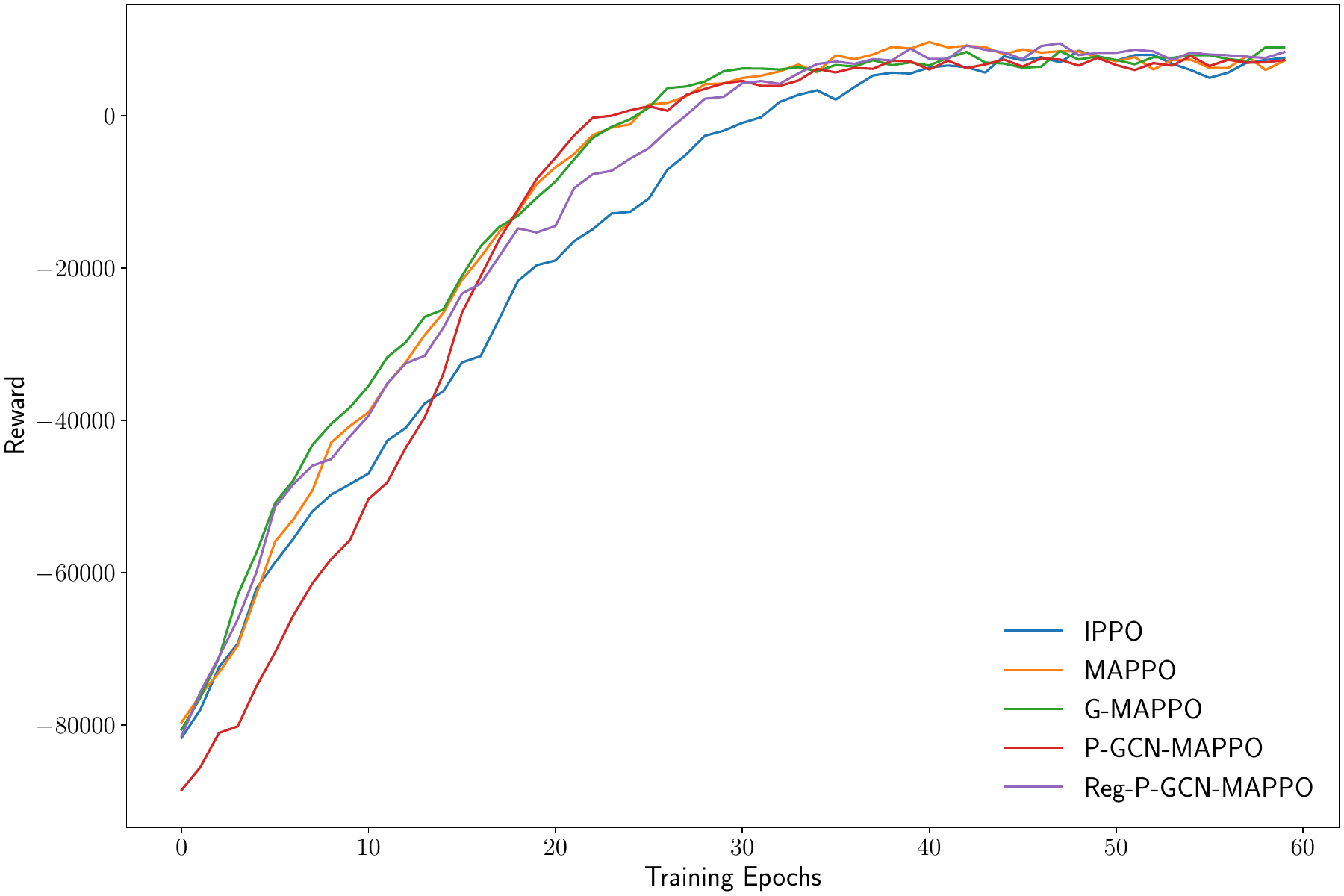}
    \caption{24 agents}
  \end{subfigure}
  \caption{Figure shows the training curves for 60 iterations for the 4 different configurations} 
  \label{fig:training}

\end{figure}

Firstly, Figure \ref{fig:training}, depicts the training curves for 60 iterations across the four different configurations consisting of 6, 12, 18 and 24 agents. As the number of agents increases, it can be seen that it takes slightly longer to reach convergence due to the increased complexity of the learning process. However, the figure also shows that the policies for all configurations converge within the 60 iterations. Figure \ref{fig:training} also shows that despite the injection of noise in the value function, this does not cause instability in training due to the inherent clipping mechanism in PPO which ensures the policy update stays within a certain limit, maintaining stability.

\begin{figure}[h!]
    \centering
    \includegraphics[width=0.8\textwidth]{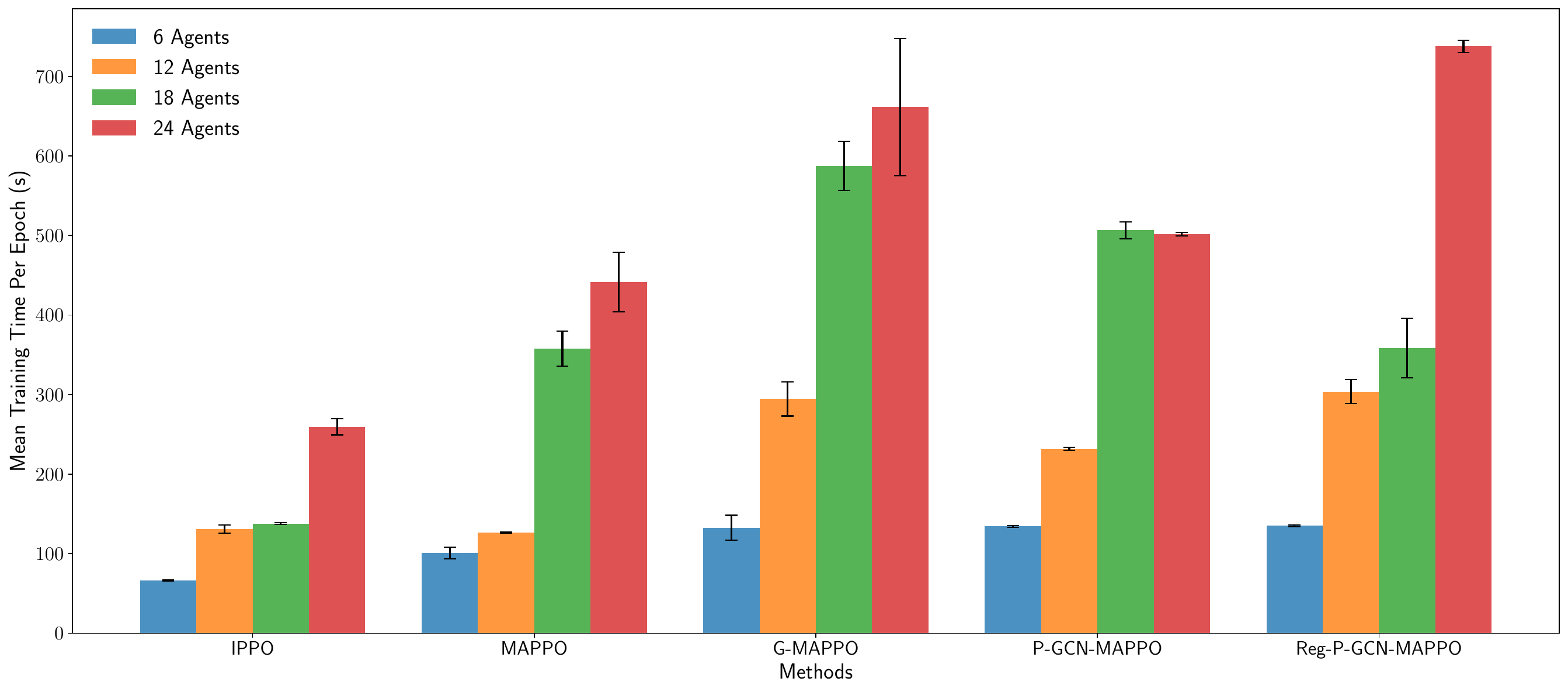}
    \caption{Box plot showing the mean training time per iteration (s) along with the associated standard deviation. Note: a z-test score was performed to remove anomalies of values which were 2 standard deviations greater than the mean value. }
    \label{fig:trainingtime}
\end{figure}

Moreover, Figure \ref{fig:trainingtime}, compares the mean training time per iteration in seconds both in terms of methodologies and number of agents. As expected, as the number of agents increases, the training time per iteration increases due to the increased complexity of the learning process. The methodologies leveraging the graph structure also have a higher mean training time per iteration because they require complex message passing between agents in the graph. G-MAPPO, P-GCN-MAPPO, Reg-P-GCN-MAPPO utilize a GNN which requires training alongside the policies. Therefore, extracting and processing information from the graph structure itself adds computational overhead compared to simpler MARL approaches. As the number of agents increases, the training time per iteration increases at a faster rate for the methodologies using GNNs as the communication and information processing becomes increasingly complex as the number of agents and complexity of the structure increases. However, Figure \ref{fig:trainingtime} shows that methods with an integrated pooling mechanism have faster training times than those without one. This further emphasizes the scalability advantage of integrating a pooling mechanism compared to a standard GNN architecture in a MARL framework. The integration of the pooling mechanism reduces the dimensionality of the input to the central critic, reducing the computational overhead which leads to faster training times. Therefore, as opposed to naïvely concatenating information, MARL frameworks that employ smart information aggregation techniques can offer several advantages: reduced computational complexity, improved scalability and potential for improved performance by efficient aggregation that can lead to better representation of the global state. 

\section{Conclusions and Future Work} \label{sec:conc}

In this work, we propose a new methodology that develops a decentralized decision-making framework for inventory management. Our framework leverages the inherent graph structure and offers several advantages that enhance the efficiency and effectiveness of inventory systems. 

Firstly, our approach redefines the action space by parameterizing a known heuristic inventory control policy which is often discontinuous in nature. Unlike traditional heuristic methods where the policy is not conditioned on the state of the system, leveraging an RL approach means the heuristic is conditioned on the state of the system. This not only allows for a more flexible decision-making framework but ensures earlier adoption of novel optimization techniques in industry due to its interpretability. 

Secondly, our methodology overcomes information sharing constraints at an online level but trains the control policies in a collaborative framework. This ensures effective coordination and communication is present within the different entities of the inventory management system. The communication between entities is further enhanced by leveraging the inherent graph structure of a supply chain. This also shifts the costs from online to offline, resulting in a more efficient online stochastic optimal control policy. The methodology results in a closed-loop solution rather than an open-loop optimization problem which enhances fast, real-time decision-making capabilities. 

In summary, as opposed to naïvely concatenating information, MARL frameworks that leverage information aggregation techniques can offer several advantages: reduced computational complexity, improved scalability and potential for improved performance by efficient aggregation that can lead to better representation of the global state. 

Future work will focus on integrating attention networks that weigh the importance of neighboring nodes which will allow the frameworks to retain more local, spatial information. The work will also be expanded to more complex systems with a larger number of products and a non-stationarity demand to mimic real-world conditions. 

The codes are available at: The codes are available at: \href{https://github.com/OptiMaL-PSE-Lab/MultiAgentRL_InventoryControl}{github.com/OptiMaL-PSE-Lab/MultiAgentRL\_InventoryControl.}

\section*{Acknowledgements}
Niki Kotecha acknowledges support from I-X at Imperial College London. 

\bibliographystyle{elsarticle-harv} 
\bibliography{ref}

\newpage

\appendix
\section{Neural Network Architectures} \label{appendix:NN}
The methodology consists of three main modules: multiple agent-specific actor networks, a centralized critic network, and a graph-based module utilizing a Graph Convolutional Network (GCN). 

The actor (policy) networks are composed of three fully connected layers. The centralized critic follows a similar structure but outputs a scalar value without a Tanh activation.The graph module employs a Graph Convolutional Network (GCN) to capture spatial dependencies between agents. The architecture details are summarized in Table~\ref{tab:nn_architectures}. 

\begin{table*}[h]
    \centering
    \resizebox{\textwidth}{!}{
    \begin{tabular}{lp{4.2cm}p{4.2cm}p{4.2cm}}
        \toprule
        \textbf{Parameter} & \textbf{Actor (Policy)} & \textbf{Critic } & \textbf{GCN} \\
        \hline
        
        Framework     & PyTorch (RLlib) & PyTorch (RLlib) & PyTorch (PyG) \\
        
        Architecture  & MLP (3 layers)  & MLP (3 layers)  & GCN (3 layers) \\
        
        Input         & State features  & State \& Action & Node features + adjacency \\
        
        Hidden Layers & 128-128-128        & 256-256-256     & 64-64-Output \\
        
        Activation    & ReLU, \textbf{tanh (output)} & ReLU & ReLU \\
        
        Output        & Action distribution (mean \& std),  & Q-value estimate & Processed node embeddings \\
        
        Optimizer     & Adam            & Adam            & Adam \\
        
        Learning Rate & \(1 \times 10^{-3}\) & \(1 \times 10^{-3}\) & \(1 \times 10^{-3}\) \\
        
        Weight Decay  & \(1 \times 10^{-4}\) & \(1 \times 10^{-4}\) & \(1 \times 10^{-4}\) \\
\bottomrule
    \end{tabular}
    }
    \caption{Neural Network Architectures for Actor, Critic, and GCN}

    \label{tab:nn_architectures}
\end{table*}

The GCN processes agent connectivity using an adjacency matrix and updates node representations through message passing. The output embeddings are integrated into the centralized critic for value estimation and influence the policy network.

\section{Hyperparameter Values} \label{appendix:hyp}
The table below summarizes the hyperparameter values for the MARL algorithms. These were kept consistent. It is worth noting these were the predefined values used in the Ray RLlib library \citep{rllib}. 

\begin{table}[h!]
  \centering
  \caption{MARL hyperparameter values.}
  \resizebox{0.6\textwidth}{!}{
    \begin{tabular}{lc}
    \hline
    \textbf{Hyperparameter} & \textbf{Value range} \\
    \hline
    Clip parameter, $\epsilon$ & 0.3 \\
    GAE parameter, $\lambda$& 1.0 \\
    Initial KL coefficient, $\beta$ & 0.2 \\
    KL target, $d_{\text{targ}}$ & 0.01 \\
    Batch Size $|\mathcal{D}|$ & 4000 \\
    Train Batch Size & 32 \\
    Epochs & 60 \\
\bottomrule
    \end{tabular}
    }%
  \label{tab:multi agent final hyperparams}%
\end{table}%
\section{Supply Chain Network Environment Parameters} \label{appendix:sc}
The tables below shows the environment parameters for the inventory management system. 

\begin{table}[htbp]
\centering
\resizebox{\textwidth}{!}{%
\begin{tabular}{@{}cccccccccc@{}}
\toprule
\textbf{Node} & \textbf{Node Costs} & \textbf{Node Prices} & \textbf{Max Inventory} & \textbf{Max Order} & \textbf{Initial Inventory} & \textbf{Target Inventory} & \textbf{Stock Costs} & \textbf{Backlog Costs} & \begin{tabular}[c]{@{}c@{}}\textbf{Connected Nodes} \\ (Downstream)\end{tabular} \\ \midrule
0    & 0.5  & 4.0 & 100 & 100 & 100 & 10 & 0.5 & 2.5 & 1, 2 \\ 
1    & 1.0  & 6.0 & 100 & 100 & 100 & 10 & 0.5 & 2.5 & 3, 4 \\ 
2    & 1.0  & 6.0 & 100 & 100 & 100 & 10 & 0.5 & 2.5 & 4, 5 \\ 
3    & 1.5  & 8.0 & 100 & 100 & 100 & 10 & 0.5 & 2.5 & None \\ 
4    & 1.5  & 8.0 & 100 & 100 & 100 & 10 & 0.5 & 2.5 & None \\ 
5    & 1.5  & 8.0 & 100 & 100 & 100 & 10 & 0.5 & 2.5 & None \\ \bottomrule
\end{tabular}%
}
\caption{Summary of 6 Supply Chain Node Parameters Including Costs, Prices, Inventory Targets, Maximum Capacities and Downstream Connectivity}
\end{table}

\begin{table}[htbp]
\centering
\resizebox{\textwidth}{!}{%
\begin{tabular}{@{}cccccccccc@{}}
\toprule
\textbf{Node} & \textbf{Node Costs} & \textbf{Node Prices} & \textbf{Max Inventory} & \textbf{Max Order} & \textbf{Initial Inventory} & \textbf{Target Inventory} & \textbf{Stock Costs} & \textbf{Backlog Costs} & \begin{tabular}[c]{@{}c@{}}\textbf{Connected Nodes} \\ (Downstream)\end{tabular} \\ \midrule
0    & 0.5  & 4.0  & 100 & 100 & 100 & 10 & 0.5 & 2.5 & 1, 2 \\ 
1    & 1.0  & 6.0  & 100 & 100 & 100 & 10 & 0.5 & 2.5 & 3, 4 \\ 
2    & 1.0  & 6.0  & 100 & 100 & 100 & 10 & 0.5 & 2.5 & 5, 6 \\ 
3    & 1.5  & 8.0  & 100 & 100 & 100 & 10 & 0.5 & 2.5 & 7, 8 \\ 
4    & 1.5  & 8.0  & 100 & 100 & 100 & 10 & 0.5 & 2.5 & 9    \\ 
5    & 1.5  & 8.0  & 100 & 100 & 100 & 10 & 0.5 & 2.5 & 10, 11 \\ 
6    & 1.5  & 8.0  & 100 & 100 & 100 & 10 & 0.5 & 2.5 & None \\ 
7    & 2.0  & 10.0 & 100 & 100 & 100 & 10 & 0.5 & 2.5 & None \\ 
8    & 2.0  & 10.0 & 100 & 100 & 100 & 10 & 0.5 & 2.5 & None \\ 
9    & 2.0  & 10.0 & 100 & 100 & 100 & 10 & 0.5 & 2.5 & 11   \\ 
10   & 2.0  & 10.0 & 100 & 100 & 100 & 10 & 0.5 & 2.5 & 11   \\ 
11   & 2.0  & 10.0 & 100 & 100 & 100 & 10 & 0.5 & 2.5 & None \\ \bottomrule
\end{tabular}%
}
\caption{Summary of 12 Supply Chain Node Parameters Including Costs, Prices, Inventory Targets, Maximum Capacities and Downstream Connectivity}
\end{table}

\begin{table}[htbp]
\centering
\resizebox{\textwidth}{!}{%
\begin{tabular}{@{}cccccccccc@{}}
\toprule
\textbf{Node} & \textbf{Node Costs} & \textbf{Node Prices} & \textbf{Max Inventory} & \textbf{Max Order} & \textbf{Initial Inventory} & \textbf{Target Inventory} & \textbf{Stock Costs} & \textbf{Backlog Costs} & \begin{tabular}[c]{@{}c@{}}\textbf{Connected Nodes} \\ (Downstream)\end{tabular} \\ \midrule
0    & 0.5  & 4.0  & 100 & 100 & 100 & 10 & 0.5 & 2.5 & 1, 2 \\ 
1    & 1.0  & 6.0  & 100 & 100 & 100 & 10 & 0.5 & 2.5 & 3, 4 \\ 
2    & 1.0  & 6.0  & 100 & 100 & 100 & 10 & 0.5 & 2.5 & 5, 6 \\ 
3    & 1.5  & 8.0  & 100 & 100 & 100 & 10 & 0.5 & 2.5 & 7, 8 \\ 
4    & 1.5  & 8.0  & 100 & 100 & 100 & 10 & 0.5 & 2.5 & 9    \\ 
5    & 1.5  & 8.0  & 100 & 100 & 100 & 10 & 0.5 & 2.5 & 10   \\ 
6    & 1.5  & 8.0  & 100 & 100 & 100 & 10 & 0.5 & 2.5 & None \\ 
7    & 2.0  & 10.0 & 100 & 100 & 100 & 10 & 0.5 & 2.5 & 11, 12, 13 \\ 
8    & 2.0  & 10.0 & 100 & 100 & 100 & 10 & 0.5 & 2.5 & 12   \\ 
9    & 2.5  & 12.0 & 100 & 100 & 100 & 10 & 0.5 & 2.5 & 14   \\ 
10   & 2.5  & 12.0 & 100 & 100 & 100 & 10 & 0.5 & 2.5 & 15   \\ 
11   & 2.5  & 12.0 & 100 & 100 & 100 & 10 & 0.5 & 2.5 & 16, 17 \\ 
12   & 2.5  & 12.0 & 100 & 100 & 100 & 10 & 0.5 & 2.5 & None \\ 
13   & 2.5  & 12.0 & 100 & 100 & 100 & 10 & 0.5 & 2.5 & 17   \\ 
14   & 2.5  & 12.0 & 100 & 100 & 100 & 10 & 0.5 & 2.5 & 17   \\ 
15   & 3.0  & 14.0 & 100 & 100 & 100 & 10 & 0.5 & 2.5 & None \\ 
16   & 3.0  & 14.0 & 100 & 100 & 100 & 10 & 0.5 & 2.5 & None \\ 
17   & 3.0  & 14.0 & 100 & 100 & 100 & 10 & 0.5 & 2.5 & None \\ \bottomrule
\end{tabular}%
}
\caption{Summary of 18 Supply Chain Node Parameters Including Costs, Prices, Inventory Targets, Maximum Capacities and Downstream Connectivity}
\end{table}

\begin{table}[htbp]
    \centering
    \resizebox{\textwidth}{!}{%
    \begin{tabular}{@{}cccccccccc@{}}
        \toprule
\textbf{Node} & \textbf{Node Costs} & \textbf{Node Prices} & \textbf{Max Inventory} & \textbf{Max Order} & \textbf{Initial Inventory} & \textbf{Target Inventory} & \textbf{Stock Costs} & \textbf{Backlog Costs} & \begin{tabular}[c]{@{}c@{}}\textbf{Connected Nodes} \\ (Downstream)\end{tabular} \\ \midrule
0    & 0.5  & 4   & 100 & 100 & 100 & 10 & 0.5 & 2.5 & 1, 2 \\ 
1    & 1.0  & 6   & 100 & 100 & 100 & 10 & 0.5 & 2.5 & 3, 4 \\ 
2    & 1.0  & 6   & 100 & 100 & 100 & 10 & 0.5 & 2.5 & 5, 6 \\ 
3    & 1.5  & 8   & 100 & 100 & 100 & 10 & 0.5 & 2.5 & 7, 8 \\ 
4    & 1.5  & 8   & 100 & 100 & 100 & 10 & 0.5 & 2.5 & 9 \\ 
5    & 1.5  & 8   & 100 & 100 & 100 & 10 & 0.5 & 2.5 & 10 \\ 
6    & 1.5  & 8   & 100 & 100 & 100 & 10 & 0.5 & 2.5 & None \\ 
7    & 2.0  & 10  & 100 & 100 & 100 & 10 & 0.5 & 2.5 & 11, 12, 13 \\ 
8    & 2.0  & 10  & 100 & 100 & 100 & 10 & 0.5 & 2.5 & 12 \\ 
9    & 2.0  & 10  & 100 & 100 & 100 & 10 & 0.5 & 2.5 & 14 \\ 
10   & 2.0  & 10  & 100 & 100 & 100 & 10 & 0.5 & 2.5 & 15 \\ 
11   & 2.5  & 12  & 100 & 100 & 100 & 10 & 0.5 & 2.5 & 16, 17 \\ 
12   & 2.5  & 12  & 100 & 100 & 100 & 10 & 0.5 & 2.5 & None \\ 
13   & 2.5  & 12  & 100 & 100 & 100 & 10 & 0.5 & 2.5 & 17 \\ 
14   & 2.5  & 12  & 100 & 100 & 100 & 10 & 0.5 & 2.5 & 17, 18 \\ 
15   & 2.5  & 12  & 100 & 100 & 100 & 10 & 0.5 & 2.5 & 19 \\ 
16   & 3.0  & 14  & 100 & 100 & 100 & 10 & 0.5 & 2.5 & 20 \\ 
17   & 3.0  & 14  & 100 & 100 & 100 & 10 & 0.5 & 2.5 & 20, 21 \\ 
18   & 3.0  & 14  & 100 & 100 & 100 & 10 & 0.5 & 2.5 & 22 \\ 
19   & 3.0  & 14  & 100 & 100 & 100 & 10 & 0.5 & 2.5 & 22, 23 \\ 
20   & 3.5  & 16  & 100 & 100 & 100 & 10 & 0.5 & 2.5 & None \\ 
21   & 3.5  & 16  & 100 & 100 & 100 & 10 & 0.5 & 2.5 & None \\ 
22   & 3.5  & 16  & 100 & 100 & 100 & 10 & 0.5 & 2.5 & None \\ 
23   & 3.5  & 16  & 100 & 100 & 100 & 10 & 0.5 & 2.5 & None \\ \bottomrule
\end{tabular}%
}
\caption{Summary of 24 Supply Chain Node Parameters Including Costs, Prices, Inventory Targets, Maximum Capacities and Downstream Connectivity}
\end{table}

\end{document}